\documentclass[useAMS]{mn2e}

\usepackage{times}
\usepackage{graphicx}
\usepackage{amsmath}
\usepackage{amssymb}
\usepackage{color}
\newcommand{\beq}{\begin{equation}}
\newcommand{\bea}{\begin{eqnarray}}
\newcommand{\eeq}{\end{equation}}
\newcommand{\eea}{\end{eqnarray}}

\title[SSC spectra at large $Y$] {The
  synchrotron-self-Compton spectrum of relativistic blast waves at
  large $Y$}

\author[M. Lemoine] {Martin
  Lemoine$^1$\thanks{e-mail:{\tt lemoine@iap.fr}}\\
  $^1$Institut d'Astrophysique de Paris, CNRS, UPMC,
  98 bis boulevard Arago, F-75014 Paris, France }

\begin{document}

\date{}

\pubyear{2008}

\maketitle

\label{firstpage}

\begin{abstract}
Recent analyses of multiwavelength light curves of gamma-ray bursts
afterglows point to values of the magnetic turbulence well below the
canonical $\sim1\,$\% of equipartition, in agreement with theoretical
expectations of a micro-turbulence generated in the shock precursor,
which then decays downstream of the shock front through collisionless
damping. As a direct consequence, the Compton parameter $Y$ can take
large values in the blast. In the presence of decaying
micro-turbulence and/or as a result of the Klein-Nishina suppression
of inverse Compton cooling, the $Y$ parameter carries a non-trivial
dependence on the electron Lorentz factor, which modifies the spectral
shape of the synchrotron and inverse Compton components. This paper
provides detailed calculations of this synchrotron-self-Compton
spectrum in this large $Y$ regime, accounting for the possibility of
decaying micro-turbulence. It calculates the expected temporal and
spectral indices $\alpha$ and $\beta$ customarily defined by
$F_\nu\,\propto\,t_{\rm obs}^{-\alpha}\nu^{-\beta}$ in various
spectral domains. This paper also makes predictions for the very high
energy photon flux; in particular, it shows that the large $Y$ regime
would imply a detection rate of gamma-ray bursts at $>10\,$GeV several
times larger than currently anticipated.
\end{abstract}


\section{Introduction}\label{sec:introd}
The afterglow radiation of gamma-ray bursts, which spans the range
from the radio domain to the X-ray and possible higher frequencies,
with a characteristic decrease in time of the peak flux, is nicely
interpreted in terms of the synchrotron radiation of
ultra-relativistic electron that have been accelerated at the forward
ultra-relativistic collisionless shock wave of the outflow, see
e.g. Piran (2004) for a detailed review.

This model thus opens a beautiful connection between astronomical data
and the microphysics of the rather extreme phenomena that relativistic
collisionless shocks represent. Zooming in on microphysical scales
reveals the shock front as a micro-turbulent magnetic barrier which
isotropizes the incoming background plasma, see e.g. Moiseev \&
Sagdeev (1963) for pioneering studies and Kato \& Takabe (2008),
Spitkovsky (2008a) for detailed numerical simulations. This
micro-turbulent barrier itself results from the build-up of
electromagnetic micro-instabilities ahead of the shock front, in the
shock precursor where supra-thermal particles mix with the incoming
background plasma, e.g. Medvedev \& Loeb (1999), Wiersma \& Achterberg
(2004), Lyubarksy \& Eichler (2006), Achterberg \& Wiersma (2007),
Achterberg et al. (2007), Bret et al. (2010), Lemoine \& Pelletier
(2010, 2011), Rabinak et al. (2011), Shaisultanov et al. (2012),
Lemoine et al. (2014a, b).

The resulting micro-turbulence, with typical length scale
$\lambda_{\delta B}\,\sim\,c/\omega_{\rm pi}$ ($\omega_{\rm pi}$ the
ion plasma frequency) and typical strength\footnote{$\epsilon_B
  \,\equiv\,\delta B^2/\left(8\mathrm{\pi}\,4\Gamma_{\rm b}^2 n m_p c^2\right)$
  represents the magnetic energy fraction of equipartition, if
  $\Gamma_{\rm b}\,\gg\,1$ represents the relative Lorentz factor
  between upstream and downstream, $n$ the upstream proper density.}
$\epsilon_B\,\sim\,0.01$, thus emerges as a key ingredient for the
microphysics of collisionless shocks. Actually, it also represents a
key requisite for the relativistic Fermi process, since this latter
can take place (in ideal conditions) only when an intense
micro-turbulence, with power on scales smaller than the gyroradius of
the accelerated particles, is able to unlock the particles off the
background magnetic field lines, by scattering them faster than a
gyration time in this background field (Lemoine et al. 2006, Niemiec
et al. 2006, Pelletier et al. 2009). In terms of magnetization $\sigma
\,=\,B^2/\left(8\mathrm{\pi}\,4\Gamma_{\rm b}^2 n m_p c^2\right)$, with $B$ the
background field expressed in the downstream frame, this condition
amounts to $\sigma\,\ll\,\epsilon_B^2$ (Lemoine \& Pelletier 2010,
2011, Lemoine et al. 2014a), a condition which is indeed satisfied for
the forward shock of gamma-ray bursts, since $\sigma\,\sim\,10^{-9}$
in the interstellar medium. This point of view has been confirmed by
particle-in-cell simulations (e.g. Spitkovsky 2008b, Martins et
al. 2009, Nishikawa et al. 2009, Sironi \& Spitkovsky 2009, 2011,
Sironi et al. 2013, Haugb\o lle 2011).

Finally, these micro-instabilities may also build the magnetized
turbulence in which the electrons eventually produce the afterglow
radiation in a synchrotron-like process (Medvedev \& Loeb 1999) --
jitter effects are expected to be weak in the conditions typical of
those shock waves -- provided it survives collisonless damping
downstream of the shock front (Gruzinov \& Waxman 1999). Recent
analyses of the damping of this Weibel-type micro-turbulence yield a
damping rate $\,\propto\,k^3$, where $k$ denotes a turbulent
wavenumber, indicating that small scales are dissipated first, but
that large scales may survive longer (Chang et al. 2008, Lemoine
2015). In turn, this implies that the turbulence strength, or
$\epsilon_B$, should decay as a power-law in (proper) time (or
distance) downstream of the shock front, with an index which itself
depends on the (unknown) micro-turbulent power spectrum at the shock
front. Interestingly, this time dependence of $\epsilon_B$ turns out
to be encoded in the multiwavelength synchrotron spectrum, since
electrons of different Lorentz factors cool at different times, hence
in regions of different magnetic field strengths (Rossi \& Rees 2003,
Derishev 2007, Lemoine 2013).

Although one cannot exclude that other external instabilities would
pollute the blast with magnetized turbulence, it is tempting to
consider that the generation of micro-turbulence could be responsible
at the same time for the formation of the shock, for the acceleration
of particles and for the radiation of these particles. It is
furthermore tempting to follow this thread to use the multiwavelength
spectrum of gamma-ray bursts as a tomograph of the magnetized
turbulence.  As a matter of fact, the recent detections of extended
emission of gamma-ray bursts in the $>100\,$MeV band by the Fermi-LAT
instrument do point to a net decay of the micro-turbulence behind the
shock front (Lemoine et al. 2013), with
$\epsilon_B\,\propto\,\left(t\omega_{\rm pi}/100\right)^{\alpha_t}$
and $-0.5\,\lesssim\,\alpha_t\,\lesssim\,-0.4$. This argument can be
recapped as follows: if the accelerated particles scatter in a
micro-turbulence, the maximal synchrotron photon energy is limited to
a few GeV at an observer time of $100$\,s (Kirk \& Reville 2010,
Plotnikov et al. 2013, Wang et al. 2013, Sironi et al. 2013); this
maximal energy scales as the square root of $\epsilon_{B+}$, i.e. the
magnitude of the turbulence in the vicinity of the shock front, and
the above value assumes $\epsilon_{B+}\,=\,0.01$; therefore, the
interpretation of this extended $>100\,$MeV emission as a synchrotron
process points to the existence of a strong micro-turbulence close to
the shock front; on the other hand, multiwavelength fits of the
afterglows for these Fermi-LAT gamma-ray bursts indicate that
low-energy photons are produced in regions of rather low
$\epsilon_{B-}$, of the order of $10^{-6}-10^{-5}$; this discrepancy
between $\epsilon_{B-}$ and $\epsilon_{B+}$ is naturally interpreted
as the decay of the micro-turbulence through collisionless damping. As
discussed in Lemoine et al. (2013), the decay rate
$\alpha_t\,\sim\,-0.5\,{\rm to}\,-0.4$ further matches the results of
a detailed numerical experiment reported in Keshet et al. (2009).

The decay of micro-turbulence has also been proposed as a possible
solution to the abnormal spectral indices observed in the prompt
emission phase (Derishev 2007), but admittedly, the physics of mildly
relativistic shock waves may well differ from that of
ultra-relativistic shock waves; in particular, the extended precursor
size in mildly relativistic shocks opens the way to other
instabilities operating on larger length scales. Although the present
considerations can also be generalized to the case of internal shocks,
all of the discussion that follows will focus on the external
ultra-relativistic shock front.

An interesting consequence of a decaying micro-turbulence, or more
generally, of a low average value of $\epsilon_{B}$ as measured in the
Fermi-LAT and other bursts (e.g. Kumar \& Barniol-Duran 2009, 2010, He
et al. 2011, Santana et al. 2014, Barniol-Duran 2014), is a large
Compton parameter $Y$, at least if one omits the influence of
Klein-Nishina effects, which actually depend on electron energy hence
on observed frequency, see below.  Recall indeed that in the standard
model, assuming that electrons cool through inverse Compton
interactions on their synchrotron spectrum in the Thomson limit
(i.e. neglecting Klein-Nishina -- KN -- effects), $Y\,\sim\,\left(\eta
\epsilon_e/\epsilon_B\right)^{1/2}$ at $Y\,\gg\,1$, with
$\eta\,\simeq\,{\rm min}\left[1,\left(\gamma_{\rm c}/\gamma_{\rm
    m}\right)^{2-p}\right]$ the cooling efficiency of electrons,
$\gamma_{\rm c}$ denoting the Lorentz factor of electrons which cool
on a dynamical timescale, $\gamma_{\rm m}$ the minimum Lorentz factor
of the injected electron power-law and $-p$ the index of this
power-law, e.g.  Panaitescu \& Kumar (2000), Sari \& Esin (2001),
Piran (2004). This $Y$ parameter also reflects the power of the
inverse Compton component relatively to the synchrotron component,
therefore the ratio of emission at very high energies to that in the
X-ray range. The possibility of a large $Y$ parameter should thus
increase the chances of observating gamma-ray bursts at very high
energies with upcoming \v{C}erenkov telescopes. The observation of
this inverse Compton component would then open a new spectral domain
with which one could study the microphysics of the turbulence in the
blast.

However, the computation of the synchrotron-self Compton (SSC)
spectrum of the blast in the large $Y$ regime is not trivial because
the shape of the synchrotron spectrum influences the cooling history
of the electrons, which itself determines the synchrotron spectral
shape. In particular, the KN suppression of the inverse Compton
cross-section may itself modify this cooling history, hence modify the
very shape of the synchrotron spectrum, see Nakar et al. (2009), Wang
et al. (2010); see also Barniol Duran \& Kumar (2011) for the
particular case of GRB090902B and Bo\v{s}jnak et al. (2009), Daigne et
al. (2011) for related discussions in the context of gamma-ray burst
prompt emission. In the case of decaying micro-turbulence, this issue
is more acute, because the synchrotron power $\nu F_\nu$ may be rising
with frequency, due to the fact that lower frequency photons are
emitted by electrons of longer cooling time, in regions of lower
magnetic field strength (Lemoine 2013). Therefore, the Lorentz
factor-dependent KN limit determines the radiation intensity on which
an electron can cool.  Finally, so far the influence of decaying
micro-turbulence on the synchrotron spectrum has been studied for a
fixed $Y$ parameter, independent of electron Lorentz factor, hence the
general shape of the SSC spectrum is not known in this physically
relevant case.

These considerations motivate the present study, which calculates the
SSC spectrum for a relativistic blast wave in the large $Y$
regime. Although the prime objective is to understand how the decaying
micro-turbulence affects the SSC spectrum, the present discussion also
addresses the case of a uniform, low value of $\epsilon_B$.  The
results are applied to the case of the GRB afterglows, by calculating
the spectrum at various observer time and by plotting, in particular,
the spectral slopes in various frequency windows. The spectrum is also
evaluated in the multi-GeV range, with a proper account of synchrotron
and SSC contributions, to make clear predictions for future
\v{C}erenkov observatories such as HAWK and CTA.

This paper is organized as follows: Sec.~\ref{sec:SSC} presents an
analytical description of the spectrum in the slow-cooling limit,
accounting for KN effects, and introduces a fast algorithm to compute
this spectrum in the fast cooling regime; Sec.~\ref{sec:lc} then
computes the light curves and plots the temporal and spectral indices
$\alpha$ and $\beta$, commonly defined by $F_\nu\,\propto\,t_{\rm
  obs}^{-\alpha}\nu^{-\beta}$, and gives predictions for the very high
energy photon flux; finally, Sec.~\ref{sec:conc} summarizes these
results and provides some conclusions.

\section{SSC spectrum}\label{sec:SSC}
\subsection{Set-up}\label{sec:SSCgen}
The set-up is as follows: electrons are swept-up by a relativistic
shock front propagating in a density profile $n\,\propto\,r^{-k}$,
then accelerated on a short time-scale to a power-law ${\rm
  d}N_{e,0}/{\rm
  d}\gamma\,\propto\,\gamma^{-p}\Theta(\gamma-\gamma_{\rm m})$ above a
minimal Lorentz factor $\gamma_{\rm m}$. The minimal Lorentz factor is
defined as usual by $\gamma_{\rm m}\,\equiv\, \epsilon_e \Gamma_{\rm
  b}m_p/m_e (p-2)/(p-1)$, with $\epsilon_e\,\sim\,0.1$. Cooling takes
place on much longer timescales.  When describing the cooling history,
the total inverse Compton cross-section is modelled as a top-hat, with
$\sigma=\sigma_{\rm T}$ for $\nu<\tilde\nu(\gamma)$ and zero
otherwise, $\tilde\nu(\gamma)$ denoting the frequency of photons with
which electrons of Lorentz factor $\gamma$ interact at the KN limit
(Nakar et al. 2009, Wang et al. 2010), i.e.
\begin{equation}
\tilde\nu\,\equiv\,\frac{\Gamma_{\rm b}m_ec^2}{h \gamma(1+z)}\ .\label{eq:tildenu}
\end{equation}
All frequencies are written in the observer rest frame; as mentioned
above, $\Gamma_{\rm b}$ denotes the Lorentz factor of the blast in the
source rest frame. The Compton parameter can then be approximated as
\begin{equation}
Y(\gamma)\,\simeq\,\frac{U_{\rm
    rad}\left[\nu<\tilde\nu(\gamma)\right]}{U_B(\gamma)}\ ,\label{eq:Y}
\end{equation}
where 
\begin{equation}
U_{\rm rad}\left[<\tilde\nu(\gamma)\right]\,=\,\int_0^{\tilde\nu(\gamma)}{\rm
      d}\nu\,U_\nu(\nu)
\end{equation}
represents the comoving radiation energy density at frequencies
$\nu<\tilde\nu(\gamma)$, on which the electron of Lorentz factor
$\gamma$ can cool. Of course, if $\tilde\nu >\nu_{\rm peak}$ at which
the differential energy density $U_{\nu}(\nu)$ reaches its maximum,
then $U_{\rm rad}\left[\nu<\tilde\nu(\gamma)\right]\,\sim\, \nu_{\rm
  peak}U_{\nu_{\rm peak}}\,\sim\,U_{\rm rad}$ and it does not depend
on $\gamma$ ($U_{\rm rad}$ denotes here the total comoving energy
density). In principle, $U_{\rm rad}$ includes all form of radiation,
synchrotron and inverse Compton alike; however, multiple Compton
scattering can be neglected for standard gamma-ray bursts parameters,
hence $U_{\rm rad}$ is hereafter determined with the synchrotron
spectrum only.

The quantity $U_B(\gamma)$ represents the energy density contained in
the magnetic field and it depends on $\gamma$ if the turbulence decays
in (proper) time behind the shock, because electrons of different
Lorentz factors then cool in regions of different magnetic field
strengths. In this work, the decay law of the turbulence takes the
power-law form $\epsilon_B(t)\,\propto\, t^{\alpha_t}$ far from the
shock front and $\epsilon_B\,\sim\,\epsilon_{B+}\,=\,0.01$ close to
the shock front. How far is expressed in terms of the proper time $t$
since the plasma element was injected through the shock,
i.e. $t\,\equiv\,x/\beta_{\rm d}$ in terms of the downstream
(comoving) distance to the shock front $x$ and $\beta_{\rm d}$, the
shock front velocity relative to the downstream rest frame. According
to PIC simulations, the characteristic (temporal) scale $\Delta$
separating far from close to the shock front is $\Delta
\,\sim\,10^{2-3}\,\,\omega_{\rm pi}$, see Chang et al. (2008), Keshet
et al. (2009) for simulations, as well as Lemoine (2013), Lemoine
(2015) for discussions of this issue; for reference, $\omega_{\rm
  pi}^{-1}\,\sim\, 7.5\times10^{-4}\,n_0^{-1/2}\,$s for a relativistic
blast wave propagating in a medium of proton density
$n_0\,$cm$^{-3}$. The uncertainty on the value of $\Delta$ does not
really influence the results presented below, because it can be
embedded in that associated to the decay power-law exponent
$\alpha_t$. Recent work mentioned above suggest
$\alpha_t\,\sim\,-0.5\rightarrow -0.4$ for a decay law $\epsilon_B
\,\sim\,\epsilon_{B+}\left(t\omega_{\rm pi}/100\right)^{\alpha_t}$,
hence the following adopts $\Delta \,=\,100\omega_{\rm pi}^{-1}$ and
$\alpha_t\,=\,-0.4$. This decay implies that the minimum magnetic
field in the blast, close to the contact discontinuity, is
characterized by the equipartition fraction
$\epsilon_{B-}\,=\,\epsilon_{B+}\left(t_{\rm
  dyn}/\Delta\right)^{\alpha_t}$ in terms of the dynamical timescale
$t_{\rm dyn}\,=\,r/\left(\Gamma_{\rm b}c\right)$.  Typical values are
$t_{\rm dyn}\,\sim\,10^5-10^6\,$s at an observer time of $10^4\,$s
(e.g. $\Gamma_{\rm b}\,\sim\,20-30$ and
$r\,\sim\,10^{17}-10^{18}\,$cm), with a mild dependence on the model
parameters, leading to values of the order of
$\epsilon_{B-}\,\sim\,10^{-5}$ for $\alpha_t\,\sim\,-0.5$. Note that
$t_{\rm dyn}/\Delta\,\propto\,t_{\rm obs}^{(5-2k)/(8-2k)}$ evolves
slowly as a function of observer time, hence so does $\epsilon_{B-}$.

In this work, it is assumed that a particle emits its synchrotron
radiation at the location at which it cools, if it cools on a
dynamical timescale; in the opposite limit, it is assumed that it
emits its synchrotron photons in the magnetic field close to the
contact discontinuity, of strength $\delta B_{-}$ (associated to the
parameter $\epsilon_{B-}$). The justification of this approximation is
as follows: the energy emitted in synchrotron photons by a particle up
to time $t$, along its trajectory downstream of the shock, can be
written as
\begin{equation}
E_{\rm syn}\,=\,\frac{1}{6\mathrm{\pi}}\sigma_{\rm T}c\int_0^t{\rm d}\tau \,\delta
B^2(\tau)\gamma_e^2(\tau)\beta_e^2(\tau)\label{eq:esyn}
\end{equation}
as a function of the time dependent Lorentz factor $\gamma_e(\tau)$ of
the particle along its cooling trajectory, with
$\beta_e(\tau)\,\sim\,1$ the particle velocity in units of $c$. One
can then show that $E_{\rm syn}$ is dominated by the contribution at
$t\,\sim\,t_{\rm cool}(\gamma)$, where $\gamma$ denotes the initial
value $\gamma_e(0)$, as follows. At early times $\tau\,\ll\,t_{\rm
  cool}(\gamma)$, $\gamma_e(\tau)\,\sim\,\gamma$ and the integrand
scales as $U_B(\tau)\,\propto\,\tau^{\alpha_t}$, hence the integral is
dominated by the large time behavior if $\alpha_t\,>\,-1$, which is
indeed an explicit assumption of the present work. At late times
$\tau\,\gg\,t_{\rm cool}(\gamma)$, one can obtain an upper bound on
how fast $\gamma_e(\tau)$ decreases by considering synchrotron losses
only, i.e. by neglecting inverse Compton losses. The integration of
\begin{equation}
\frac{{\rm d}\gamma_{e}}{{\rm d}\tau}\,=\,-\frac{4}{3}\frac{\sigma_{\rm
  T}cU_B(\tau)}{m_ec^2}\gamma_{e}^2(\tau)\beta_{e}(\tau)^2
\end{equation}
leads to $\gamma_e(\tau)\,\sim\,\gamma\,\left[\tau/t_{\rm
    cool}(\gamma)\right]^{-1-\alpha_t}$ for $\tau\,\gg\,t_{\rm
  cool}(\gamma)$. Therefore, the integrand in Eq.~(\ref{eq:esyn})
behaves as $\tau^{-2-\alpha_t}$ for $\tau\,>\,t_{\rm cool}(\gamma)$,
hence $E_{\rm syn}$ is indeed dominated by the contribution at
$t\,\sim\,t_{\rm cool}(\gamma)$ provided $\alpha_t\,>\,-1$.

This thus supports the above approximation that a particle with
initial Lorentz factor $\gamma\,>\,\gamma_{\rm c}$ emits its
synchrotron radiation on a magnetic field of strength $\delta
B\left[t_{\rm cool}(\gamma)\right]$, with $t_{\rm cool}(\gamma)$ the
cooling time of the particle. Of course, if $t_{\rm
  cool}(\gamma)\,>\,t_{\rm dyn}$, the particle does not actually cool,
and $E_{\rm syn}$ is then dominated by the upper bound
$t\,\sim\,t_{\rm dyn}$, i.e. particles radiate synchrotron radiation
in the relaxed magnetic field at the back of the blast, of strength
$\delta B_-$.

Since $t_{\rm cool}(\gamma_{\rm c})=t_{\rm dyn}$ by definition, and
since $\delta B(t_{\rm dyn})=\delta B_-$, also by definition,  
\begin{equation}
U_B(\gamma)\,\simeq\,U_{B-}\,{\rm max}\left[1,\left(\frac{t_{\rm
      cool}(\gamma)}{t_{\rm
      dyn}}\right)^{\alpha_t}\right] \label{eq:UB}
\end{equation}
with $U_{B-}\,\equiv\,\delta B_{-}^2/(8\mathrm{\pi})\,=\,U_B(\gamma_{\rm
  c})$. 

In line with the above discussion, electrons of Lorentz factor
$\gamma$ radiate their energy through synchrotron at a typical frequency
\begin{equation}
\nu_{\rm syn}(\gamma)\,\propto\,\delta B\left[t_{\rm
    cool}(\gamma)\right]\gamma^2\,\simeq\,\nu_{\rm c}
\left[\frac{t_{\rm cool}(\gamma)}{t_{\rm
      dyn}}\right]^{\alpha_t/2}\left(\frac{\gamma}{\gamma_{\rm
    c}}\right)^2 \ .\label{eq:nup}
\end{equation}
Here, $\nu_{\rm c}\,\equiv\,\nu_{\rm syn}(\gamma_{\rm c})$ denotes the
synchrotron peak frequency for particles of Lorentz factor
$\gamma_{\rm c}$.

Finally, the cooling timescale can be written
\begin{eqnarray}
t_{\rm cool}(\gamma)&\,\simeq\,&t_{\rm
  dyn}\,\frac{1+Y_c}{1+Y(\gamma)}\frac{U_{B-}}{U_B(\gamma)}\frac{\gamma_{\rm
  c}}{\gamma}\ ,\nonumber\\ 
&\,\simeq\,& t_{\rm
  dyn}\,\left[\frac{1+Y(\gamma)}{1+Y_{\rm
      c}}\right]^{-1/(1+\alpha_t)}\,\left( \frac{\gamma}{\gamma_{\rm
    c}}\right)^{-1/(1+\alpha_t)}\ .\label{eq:tcool}
\end{eqnarray}
The second equality is obtained by replacing $U_B(\gamma)/U_{B-}$ with
its value given in Eq.~(\ref{eq:UB}), assuming $\gamma\,>\,\gamma_{\rm
  c}$.

Equations~(\ref{eq:Y}), (\ref{eq:UB}), (\ref{eq:nup}) and
(\ref{eq:tcool}) then allow to derive the following scalings for
$\gamma\,>\,\gamma_{\rm c}$:
\begin{equation}
\frac{U_B(\gamma)}{U_{B-}}\,\simeq\,\left[\frac{1+Y(\gamma)}{1+Y_{\rm
      c}}\right]^{-\alpha_t/(1+\alpha_t)}\left(\frac{\gamma}{\gamma_{\rm
      c}}\right)^{-\alpha_t/(1+\alpha_t)}\ ,\label{eq:UBscal}
\end{equation}
\begin{equation}
\frac{\nu_{\rm syn}(\gamma)}{\nu_{\rm
    c}}\,\simeq\,\left[\frac{1+Y(\gamma)}{1+Y_{\rm
      c}}\right]^{-\alpha_t/[2(1+\alpha_t)]}\left(\frac{\gamma}{\gamma_{\rm
      c}}\right)^{2-\alpha_t/[2(1+\alpha_t)]}\ ,\label{eq:nuscal}
\end{equation}
\begin{equation}
\frac{Y(\gamma)\left[1+Y(\gamma)\right]^{-\alpha_t/(1+\alpha_t)}}{Y_{\rm
    c}(1+Y_{\rm c})^{-\alpha_t/(1+\alpha_t)}}\,\simeq\,\frac{U_{\rm rad}(\nu<\tilde\nu)}{U_{\rm rad}(\nu<\tilde\nu_{\rm c})}\left(\frac{\gamma}{\gamma_{\rm
      c}}\right)^{\alpha_t/[(1+\alpha_t)]}\ .\label{eq:Yscal}
\end{equation}
$Y_{\rm c}$ stands for $Y(\gamma_{\rm c})$.

Once the ratio $U_{\rm rad}(\nu<\tilde\nu)/U_{\rm
  rad}(\nu<\tilde\nu_{\rm c})$ -- which captures KN effects -- has
been specified, it is possible to derive the scalings of $Y(\gamma)$
as a function of $\gamma$, hence of $\nu_{\rm syn}(\gamma)$. One
should emphasize that in the above expressions, $\gamma$ represents
the initial Lorentz factor of the particle, after acceleration has
shaped the power-law, but before cooling has effectively taken place.

\subsection{Synchrotron spectrum}\label{sec:gen-syn}
Standard calculations of the synchrotron spectrum of a blast wave
generally derive the stationary electron distribution in the blast, by
solving a transport equation in momentum space, averaged over the
depth of the blast, accounting for injection of the power-law at the
shock and for cooling downstream (e.g. Sari et al. 1998, Sari \& Esin
2001); this yields a standard broken power-law shape ${\rm d}N_e/{\rm
  d}\gamma_e$ with indices $-2$ for $\gamma_{\rm
  c}\,<\,\gamma\,<\,\gamma_{\rm m}$ or $-p$ for $\gamma_{\rm
  m}\,<\,\gamma\,<\,\gamma_{\rm c}$ and $-p-1$ for ${\rm
  max}(\gamma_{\rm c},\gamma_{\rm m})\,<\,\gamma$. However, one can
also derive the synchrotron spectrum by a direct mapping of the energy
initially stored into the electron population to that radiated in
synchrotron; this approach is simpler in the present case and it works
as follows.
  
Electrons injected with Lorentz factor $\gamma\,>\,{\rm
  max}(\gamma_{\rm m},\gamma_{\rm c})$ emit a fraction
$\left[1+Y(\gamma)\right]^{-1}$ of their energy in synchrotron
radiation; the energy density contained in such electrons (within an
interval ${\rm d}\ln\gamma$) is itself a fraction
$(p-2)\left(\gamma/\gamma_{\rm m}\right)^{2-p}\,{\rm d}\ln\gamma$ of
the energy density $U_e$ contained in electrons immediately behind the
shock front, before cooling has started to take place;
$U_e\,=\epsilon_e\,4\Gamma_{\rm b}^2 nm_p c^2$. Consequently, the
synchrotron flux received at frequency $\nu=\nu_{\rm syn}(\gamma)$ can
be written:
\begin{equation}
\nu F_{\nu,\rm syn}\,\simeq\, \frac{1}{4\mathrm{\pi} D_{\rm
    L}^2}\frac{4}{3}\Gamma_{\rm b}^2 \frac{4\mathrm{\pi} r^2 c (p-2)
  U_e}{1+Y(\gamma)}\, \left(\frac{\gamma}{\gamma_{\rm
    m}}\right)^{2-p}\frac{{\rm d}\ln\gamma}{{\rm
    d}\ln\nu} \label{eq:nuFnuhi}
\end{equation}
The above assumes $p\,>\,2$, but it can be generalized to any index
$p\,>\,1$; indeed, the above picture assumes that particles with
$\gamma\,>\,{\rm max}(\gamma_{\rm m},\gamma_{\rm c})$ radiate their
energy at $\nu_{\rm syn}(\gamma)$ then do not contribute anymore to
the synchrotron spectrum, which is a reasonable approximation if
$p\,>\,1$; in contrast, if $p\,<\,1$, the radiation of the high
energy particles during their complete cooling history dominates that
of the lower energy ones.

Similarly, one can write down the flux for $\gamma_{\rm
  m}\,>\,\gamma\,>\,\gamma_{\rm c}$ or $\gamma_{\rm
  c}\,>\,\gamma\,>\,\gamma_{\rm m}$, whichever occurs, as follows:
\begin{equation}
\nu F_{\nu,\rm syn} \,\simeq\, \nu_{\rm m} F_{\nu_{\rm m},\rm
  syn} \begin{cases}\displaystyle{ \frac{\gamma}{\gamma_{\rm
        m}}\frac{1+Y_m}{1+Y}}& \quad (\gamma_{\rm
    m}\,>\,\gamma\,>\,\gamma_{\rm
    c})\\ \displaystyle{\left(\frac{\nu}{\nu_{\rm
        m}}\right)^{(3-p)/2}}& \quad (\gamma_{\rm
    c}\,>\,\gamma\,>\,\gamma_{\rm m})\label{eq:nuFnulo}\end{cases}
\end{equation}
with $Y_m\,\equiv\,Y(\gamma_m)$ and $\nu_{\rm m}\,=\,\nu_{\rm
  syn}(\gamma_{\rm m})$.  The spectrum at $\nu\,<\,\nu_{\rm c}$ is
indeed unchanged with respect to the standard case, although the value
of $U_{B-}$ must be used to compute the characteristic
frequencies. Regarding the fast cooling limit $\gamma_{\rm
  c}\,<\,\gamma\,<\,\gamma_{\rm m}$, the factor can be understood by
noting that all injected particles with $\gamma\,>\,\gamma_{\rm m}$
shift from $\gamma$ to $\gamma_{\rm c}$ during their cooling history,
and that at each point along this cooling trajectory they radiate (in
synchrotron) a fraction $\left(\gamma/\gamma_{\rm
  m}\right)\left(1+Y_{\rm m}\right)/\left(1+Y\right)$ of the energy
radiated (in synchrotron) by a particle of Lorentz factor $\gamma_{\rm
  m}$. Note also that in the limits $\alpha_t\,\rightarrow\,0$,
$Y(\gamma)\,\ll\,1$, one recovers the usual scaling $\nu F_{\nu,\rm
  syn}\,\propto\,\nu^{1/2}$ since $\nu\,\propto\,\gamma^{2}$.

Provided $\gamma_{\rm c}$ and $Y_{\rm c}$ are given, one can derive
$\nu_{\rm syn}(\gamma)$ and $Y(\gamma)$, hence $\gamma(\nu)$ and
$Y(\nu)$ by using Eqs.~(\ref{eq:UBscal}), (\ref{eq:nuscal}) and
(\ref{eq:Yscal}), then $\nu F_{\nu,\rm syn}$ using
Eqs.~(\ref{eq:nuFnuhi}) and (\ref{eq:nuFnulo}). Consider as an example
the case $Y\,\gg\,1$, omitting KN effects,
i.e. $\tilde\nu\,\rightarrow\,+\infty$ for all $\nu$: for
$\gamma>\gamma_{\rm c}$, one derives
$Y(\gamma)\,\propto\,\gamma^{\alpha_t}$ from Eq.~(\ref{eq:Yscal}),
hence $\nu_{\rm p}\,\propto\,\gamma^{2-\alpha_t/2}$ from
Eq.~(\ref{eq:nuscal}), hence $\nu F_{\nu,\rm syn} \,\propto\,
\gamma^{2-p}/(1+Y)\,\propto\, \nu^{(2-p-\alpha_t)/(2-\alpha_t/2)}$
from Eq.~(\ref{eq:nuFnuhi}). This latter scaling matches the explicit
calculation of Lemoine (2013), which computes the integrated cooling
history of the electron power-law in the decaying magnetic field. This
example also confirms that $\nu F_{\nu,\rm syn}$ rises with $\nu$ if
$\alpha_t\,\lesssim\,2-p$, hence KN effects should not be omitted in
an accurate calculation of $\nu F_{\nu,\rm syn}$.

The above provides the tools needed to calculate the synchrotron
spectrum; explicit calculations in both the slow and the fast cooling
regimes are provided in the following sections. 

\subsection{Inverse Compton spectrum}\label{sec:gen-ic}
In order to calculate the inverse Compton emissivity, the synchrotron
spectral density $F_{\nu,\rm syn}$ must be folded over the Compton
cross-section and the cooled particle distribution, integrated over
the blast.  The following derivation of the cooled distribution
function relies on the observations that ${\rm d}t_{\rm
  cool}(\gamma)/{\rm d}\gamma\,<\,0$ and that the cooling history
$\gamma_{\rm cool}(t)$ at times $t\,>\,t_0$ does not depend on the
history at times $t\,<\,t_0$, i.e. electrons of various Lorentz factor
follow a universal cooling trajectory $\gamma_{\rm cool}(t)$.

One writes ${\rm d}\dot N_{e,0}$ the (comoving) rate at which
electrons are swept-up and accelerated into a power-law in a Lorentz
factor interval ${\rm d}\gamma$. One also defines ${\rm d}N_e$, which
represents the total number of electrons in the blast in a Lorentz
factor interval ${\rm d}\gamma_e$; note the difference of notation:
$\gamma_e$ refers to a value of the Lorentz factor of the blast
averaged distribution, while $\gamma$ corresponds to the Lorentz
factor of the injection distribution, i.e. before cooling has
occurred.

The injection of particles with Lorentz factor $\gamma$ populates the
downstream with particles of Lorentz factor $\gamma_e\,=\,\gamma$ over
a fraction $t_{\rm cool}/t_{\rm dyn}$ of the depth of the blast. In
contrast, particles injected with $\gamma_{\rm
  c}\,>\,\gamma\,>\,\gamma_{\rm m}$ retain their Lorentz factor (no
cooling) and populate the whole blast. Therefore, one can write the
average distribution for $\gamma_e\,>\,\gamma_{\rm m}$ as:
\begin{equation}
\frac{{\rm d}N_e}{{\rm d}\gamma_e}\,\simeq\,\frac{{\rm d}\dot
  N_{e,0}}{{\rm d}\gamma}(\gamma\,=\,\gamma_e)\,\, {\rm
  min}\left[t_{\rm dyn},t_{\rm cool}(\gamma)\right]\quad (\gamma>\gamma_{\rm m})\ .
\end{equation}
Note also that $t_{\rm dyn}$ sets the scale for adiabatic losses, while
$t_{\rm cool}$ sets the scale for radiative losses. 

In the fast cooling regime, all electrons injected with Lorentz factor
$\gamma\,>\,\gamma_{\rm m}$ cool down to $\gamma_{\rm c}$, following a
same cooling history $\gamma_{\rm cool}(\tau)$, therefore
\begin{eqnarray}
\frac{{\rm d}N_e}{{\rm d}\gamma_e}&\,\simeq\,&\int_{0}^{t_{\rm dyn}}{\rm d}\tau\,\int_{\gamma_{\rm m}}^{+\infty}{\rm d}\gamma\,\frac{{\rm d}\dot N_{e,0}}{{\rm d}\gamma}\,
\delta\left[\gamma_e - \gamma_{\rm cool}(\tau)\right]\nonumber\\
&\,\simeq\,& \dot N_{e,0}\frac{t_{\rm cool}(\gamma_e)}{\gamma_e}\label{eq:dNfast}
\end{eqnarray}
where $\dot N_{e,0}\,=\,\int_{\gamma_{\rm m}}^{+\infty}{\rm
  d}\gamma\,{\rm d}\dot N_{e,0}/{\rm d}\gamma$ represents the total
injection rate of electrons. The last expression in
Eq.~(\ref{eq:dNfast}) follows from the definition $t_{\rm
  cool}(\gamma_e)\,=\,\gamma_e \left\vert{\rm d}\gamma_e/{\rm
  d}t\right\vert^{-1}$ after interchanging the integrals.

Using the scaling of $t_{\rm cool}(\gamma)$, one then derives
\begin{eqnarray}
\frac{{\rm d}N_e}{{\rm d}\gamma_e}&\,\simeq\,&\frac{\dot N_{e,0}t_{\rm dyn}}{\gamma_{\rm m}}
\left(\frac{\gamma_e}{\gamma_{\rm m}}\right)^{-p} \quad\quad \left(\gamma_{\rm m}<\gamma_e<\gamma_{\rm c}\right)\nonumber\\
\frac{{\rm d}N_e}{{\rm d}\gamma_e}&\,\simeq\,&\frac{\dot N_{e,0}t_{\rm dyn}}{\gamma_{\rm m}}\left(\frac{\gamma_e}{\gamma_{\rm m}}\right)^{-1}
\left(\frac{\gamma_e}{\gamma_{\rm c}}\right)^{-1/(1+\alpha_t)}\nonumber\\ &&\quad\times
\left[\frac{1+Y(\gamma_e)}{1+Y_{\rm c}}\right]^{-1/(1+\alpha_t)}\quad\quad \left(\gamma_{\rm c}<\gamma_e<\gamma_{\rm m}\right)\nonumber\\
\frac{{\rm d}N_e}{{\rm d}\gamma_e}&\,\simeq\,&\frac{\dot N_{e,0}t_{\rm dyn}}{\gamma_{\rm m}}\left(\frac{\gamma_e}{\gamma_{\rm m}}\right)^{-p}
\left(\frac{\gamma_e}{\gamma_{\rm c}}\right)^{-1/(1+\alpha_t)}\nonumber\\ &&\quad\times
\left[\frac{1+Y(\gamma_e)}{1+Y_{\rm c}}\right]^{-1/(1+\alpha_t)}\quad \quad \left[{\rm max}(\gamma_{\rm c},\gamma_{\rm m})<\gamma_e\right]\nonumber\\
&& \label{eq:dNe}
\end{eqnarray}
In the limit $\alpha_t\,\rightarrow\,0$, one recovers ${\rm d}N_e/{\rm
  d}\gamma_e\,\propto\,\left[1+Y(\gamma_e)\right]^{-1}\gamma_e^{-2}$
for $\gamma_{\rm c}\,<\,\gamma_e\,<\,\gamma_{\rm m}$ or ${\rm
  d}N_e/{\rm
  d}\gamma_e\,\propto\,\left[1+Y(\gamma_e)\right]^{-1}\gamma^{-p-1}$
for $\gamma_{\rm m}\,<\,\gamma_{\rm c}\,<\,\gamma_e$; both match the
expressions derived in Nakar et al. (2009) and in Wang et al. (2010)
for a homogeneous (non-decaying) turbulence.

It is also instructive to show that the above average electron
distribution, when associated with the appropriate (i.e. Lorentz
factor dependent) magnetic field, leads to the same scalings for the
synchrotron spectrum as Eqs.~(\ref{eq:nuFnuhi}) and
(\ref{eq:nuFnulo}). Consider for instance the regime $\gamma\,>\,{\rm
  max}\left(\gamma_{\rm c},\gamma_{\rm m}\right)$: the average
synchrotron power of the blast scales as $\nu F_{\nu,\rm
  syn}\,\propto\, U_B(\gamma_e)\gamma_e^2\,{\rm d}N_e/{\rm
  d}\ln\gamma_e$; using the scaling for $U_B(\gamma_e)$ given in
Eq.~(\ref{eq:UBscal}), one recovers $\nu F_{\nu,\rm syn}\,\propto
\left[1+Y(\gamma_e)\right]^{-1}\gamma_e^{2-p}$ indicated in
Eq.~(\ref{eq:nuFnuhi}). One can proceed similarly for the other two
regimes, noting in particular that for $\gamma_{\rm
  m}\,<\,\gamma_e\,<\,\gamma_{\rm c}$ (slow cooling), $U_B\,=\,U_{B-}$
and it no longer depends on $\gamma_e$.

Finally, using the above electron distribution, one can compute the
inverse Compton component using standard formulae:
\begin{eqnarray}
F_{\nu,{\rm IC}}(\nu_{\rm IC})&\,\simeq\,&\tau_{\rm IC}\,\int{\rm
  d}\gamma_e\,\frac{1}{N_e}\frac{{\rm d}N_e}{{\rm
    d}\gamma_e}\,\int_{0}^{1}{\rm d}q \,(1-u)g_{\rm
  KN}(q)\nonumber\\ &&\quad \times\,F_{\nu,\rm
  syn}\left[\frac{\nu_{\rm IC}}{4\gamma^2 q(1-u)}\right]
\Theta\left(1-u\right)\ ,\nonumber\\ &&\label{eq:fIC}
\end{eqnarray}
with $u\,\equiv\,(1+z)h\nu_{\rm IC}/\left(\Gamma_{\rm b}\gamma_e m_e
c^2\right)$; the function $g_{\rm KN}(q)\,=\,\left[2q\ln q +
  (1+2q)(1-q)+ G^2(1-q)/\left[2(1+G)\right]\right]$, with $G
\,\equiv\,u/(1-u)$ characterizes the energy dependence of the
Klein-Nishina cross-section, see Blumenthal \& Gould (1970) for
details.

The prefactor $\tau_{\rm IC}$ defines the optical depth to Compton
scattering; an explicit calculation leads to
\begin{equation}
\tau_{\rm IC}\,=\,3\sigma_{\rm T}\frac{N_e}{4\mathrm{\pi}
  r^2}\,=\,12\sigma_{\rm T}\Gamma_{\rm b}n ct_{\rm dyn}
\end{equation}
Note that this expression uses $N_e\,=\,4\mathrm{\pi} r^2\,c t_{\rm
  dyn}\,4\Gamma_{\rm b}n$, i.e. it only includes the electrons that
have been swept up in the last dynamical timescale, because electrons
injected at earlier times have been adiabatically cooled to Lorentz
factors $<\,\gamma_{\rm c}$ and therefore do not participate in
shaping the inverse Compton spectrum. With the above value of
$\tau_{\rm IC}$, one can verify that the energy density of the
radiation associated to the inverse Compton component is correctly
normalized to $Y_{\rm c}$ times that contained in the synchrotron
component, as can be verified by calculating $\int{\rm d}\nu_{\rm IC}
F_{\nu_{\rm IC},\rm IC}$ in terms of $\int{\rm d}\nu_{\rm
  syn}F_{\nu_{\rm syn},\rm syn}$ (neglecting the influence of
Klein-Nishina effects).

\subsection{Slow cooling}
\subsubsection{General procedure}
As discussed in detail in Nakar et al. (2009) and Wang et al. (2010),
the cooling Lorentz factors and Compton parameters $\gamma_{\rm c}$
and $Y_{\rm c}$ can be obtained in the slow cooling regime
$\gamma_{\rm m}\,<\,\gamma_{\rm c}$ from the system
\begin{eqnarray}
Y_{\rm c}\left(1+Y_{\rm
  c}\right)&\,\simeq\,&\displaystyle{\frac{\epsilon_e}{\epsilon_{B-}}\left(\frac{\gamma_{\rm
      c}}{\gamma_{\rm m}}\right)^{2-p}\,\frac{U_{\rm
      syn}(<\tilde\nu_{\rm c})}{U_{\rm syn}(<\nu_{\rm
      c})}}\nonumber\\ \gamma_{\rm
  c}&\,\simeq\,&\displaystyle{\frac{\gamma_{\rm c,syn}}{1+Y_{\rm
      c}}} \label{eq:Ycgcslow}
\end{eqnarray}
with $\gamma_{\rm c,syn}\,\equiv\,(3/4) m_e c/(\sigma_{\rm T} U_{B-}
t_{\rm dyn})$ the cooling Lorentz factor in the relaxed turbulence, in
the absence of inverse Compton losses.

Write $C_{\rm c}\,\equiv\,U_{\rm syn}(<\tilde\nu_{\rm c})/U_{\rm
  syn}(\nu_{\rm c})$ the term entering the first equation. If
$\tilde\nu_{\rm c}\,<\,\nu_{\rm c}$, $C_{\rm c}\,<\,1$ because the
peak of the synchrotron flux lies at $\nu_{\rm c}$ or above (see
below); in this limit, Klein-Nishina suppression inhibits the cooling
of $\gamma_{\rm c}$ electrons. Assuming that such electrons cool by
interacting with the $\nu_{\rm min}\,<\,\nu\,<\,\nu_{\rm c}$ part of
the synchrotron spectrum, which is generically the case, one can write
$C_{\rm c}\,=\,(\tilde\nu_{\rm c}/\nu_{\rm
  c})^{(3-p)/2}\,\propto\,\gamma_{\rm c}^{-3(3-p)/2}$ and the system
can be solved easily.

In the opposite limit, $\tilde\nu_{\rm c}\,>\,\nu_{\rm c}$; one may
have $C_{\rm c}\,\sim\,1$ if the peak of the synchrotron flux is
located at $\nu_{\rm c}$, or $C_{\rm c}\,\gtrsim\,1$ if the peak lies
at higher frequencies. In this latter case, $C_{\rm
  c}\,\simeq\,(\tilde\nu_{\rm c}/\nu_{\rm c})^{1-\beta}$, where
$\beta$ is the synchrotron spectral index defined by $F_{\nu,\rm
  syn}\,\propto\,\nu^{-\beta}$ in the spectral range above $\nu_{\rm
  c}$, calculated thereafter. The analytical calculation proposed here
uses Eq.~(\ref{eq:slowhathatnuc}) below to derive this $\beta$, but
the result depends little on this choice, because in all cases
considered, $1-\beta$ is small, meaning that the peak flux is not very
different from the flux $\nu F_\nu$ at $\nu_{\rm c}$.

The next step is to determine the critical Lorentz factors
$\widehat\gamma_{\rm c}\,\equiv\,\widehat\gamma(\nu_{\rm c})$ and
$\widehat\gamma_{\rm m}\,\equiv\,\widehat\gamma(\nu_{\rm m})$ with the general
definition (Nakar et al. 2009):
\begin{equation}
\widehat\gamma(\nu)\,=\,\frac{\Gamma_{\rm b}m_e
  c^2}{(1+z)h\nu}\ ,\label{eq:hatg}
\end{equation}
which corresponds to the Lorentz factor for which electrons interact
with photons of frequency $\nu$ at the onset of the Klein-Nishina
regime. Note that $\nu_{\rm m}$ and $\nu_{\rm c}$ are to be calculated
in the relaxed turbulence of strength $\delta B_{-}$ in this slow
cooling regime. Using Eqs.~(\ref{eq:UBscal}), (\ref{eq:nuscal}) and
(\ref{eq:Yscal}), one may then calculate the Compton parameters
$Y(\widehat\gamma_{\rm c})$ and $Y(\widehat\gamma_{\rm m})$. Note also
that the slow cooling limit $\gamma_{\rm m}<\gamma_{\rm c}$ implies
$\widehat\gamma_{\rm c}<\widehat\gamma_{\rm m}$.

Another critical Lorentz factor is $\gamma_0$, for which
$Y(\gamma_0)=1$. If $Y(\widehat\gamma_{\rm m})>1$, then $\widehat\gamma_{\rm
  m}<\gamma_0$, and $\gamma_0$ can be obtained by solving 
\begin{eqnarray}
\frac{Y(\gamma_0)\left[1+Y(\gamma_0)\right]^{-\alpha_t/(1+\alpha_t)}}{
Y(\widehat\gamma_{\rm
  m})\left[1+Y(\widehat\gamma_{\rm m})\right]^{-\alpha_t/(1+\alpha_t)}}=\left[
\frac{\tilde\nu(\gamma_0)}{\nu_{\rm m}}\right]^{4/3}
\left(\frac{\gamma_0}{\gamma_{\rm m}}\right)^{\alpha_t/(1+\alpha_t)}\nonumber\\
&&
\label{eq:g0slow}
\end{eqnarray}
which derives from Eq.~(\ref{eq:Yscal}). This latter equation assumes
that $\tilde\nu(\gamma_0)$ lies in the spectral range between
$\nu_{\rm a}$ (synchrotron self-absorption frequency) and $\nu_{\rm
  m}$, in general a very good approximation; it can be generalized to
other cases without difficulty. If $Y(\widehat\gamma_{\rm m})<1$, then
$\gamma_0<\widehat\gamma_{\rm m}$; one can repeat the above exercise
to derive $\gamma_0$, replacing the $4/3$ exponents with $(3-p)/2$,
which characterizes the spectral dependence of $\nu F_{\nu,\rm syn}$
between $\nu_{\rm m}$ and $\nu_{\rm c}$.

For improved accuracy, one may also determine the next order critical
Lorentz factor $\widehat{\widehat{\gamma}}_{\rm
  c}\,\equiv\,\Gamma_{\rm b}m_e c^2/\left[(1+z)h\nu_{\rm
    syn}(\widehat\gamma_{\rm c})\right]$. Following a procedure
similar to the above, one can derive
$\nu(\widehat{\widehat\gamma}_{\rm c})$ and
$Y(\widehat{\widehat{\gamma}}_{\rm c})$.

The following makes use of the short-hand notation:
$\widehat\nu\,=\,\nu_{\rm syn}\left(\widehat\gamma\right)$ for various
critical Lorentz factors; similarly, $\nu_{\rm syn}(\gamma_0)$ is
written $\nu_0$ for commodity.

One can derive the power-law $1-\beta'$ index of $\nu F_{\nu,\rm syn}$
as follows. Below $\nu_{\rm c}$, the scaling remains unchanged
compared to the standard case because the cooling history does not
influence the synchrotron spectrum, and the results will not be
repeated here. Above $\nu_{\rm c}$, the spectral slope is determined
by the scalings of $Y(\gamma)$, $\nu(\gamma)$ and $\nu F_{\nu,\rm
  syn}\,\propto\, \left(1+Y\right)^{-1}\gamma^{2-p}$
[Eq.~(\ref{eq:nuFnuhi})]. In particular, if $\tilde\nu(\gamma)$ lies
in a spectral domain in which $\nu F_{\nu,\rm
  syn}\,\propto\,\nu^{1-\beta}$, below the peak of the synchrotron
energy flux, then electrons of Lorentz factor $\gamma$ cool by inverse
Compton interactions with that portion of the synchrotron spectrum if
$Y(\gamma)\,\gg\,1$, in which case Eqs.~(\ref{eq:UBscal}),
(\ref{eq:nuscal}) and (\ref{eq:Yscal}) lead to
\begin{eqnarray}
Y&\,\propto\,&\gamma^{\alpha_t-(1-\beta)(1+\alpha_t)}\nonumber\\
\nu&\,\propto\,&\gamma^{2-\beta\alpha_t/2}\nonumber\\
t_{\rm cool}&\,\propto\,& \gamma^{-\beta}\label{eq:Ynuslow}
\end{eqnarray}
Assuming $\nu F_{\nu,\rm syn}\,\propto\,\nu^{1-\beta'}$ in the range
of interest around $\nu(\gamma)$, and using the above scalings, one
obtains directly
\begin{equation}
1-\beta'\,=\,\frac{2-p-\alpha_t+(1-\beta)(1+\alpha_t)}{2-\alpha_t/2+(1-\beta)\alpha_t/2}\ .
\label{eq:bpslow}
\end{equation}
The limit $\beta\rightarrow 1$ recovers the case in which $\tilde\nu$
lies above the peak of the synchrotron flux, discussed previously.

\subsubsection{Power-law segments}
As mentioned above, the synchrotron spectrum at $\nu\,<\,\nu_{\rm c}$
remains unaffected and it is not discussed here. At the upper end of
the spectral range, i.e. $\nu_0\,<\,\nu$, $Y(\gamma)\,<\,1$
because of the Klein-Nishina suppression of electron cooling, which
implies that $\nu F_{\nu,\rm syn}\,\propto\nu^{1-\beta'}$ with
\begin{equation}
1-\beta' \, =\, \frac{2-p}{2-\alpha_t/[2(1+\alpha_t)]}\quad
\left[\nu_0\,<\,\nu\right]\label{eq:slowsupnu0}
\end{equation}
as can be derived from Eq.~(\ref{eq:nuFnuhi}) with
$1+Y\,\simeq\,1$. This scaling also matches that derived by a full
computation of the electron cooling history with negligible inverse
Compton losses in Lemoine (2013). For $\alpha_t\,=\,0$, one recovers
of course the standard fast cooling index $\beta'\,=\,p/2$.

In the range $\widehat\nu_{\rm m}\,<\,\nu\,<\,\nu_0$, if it exists,
the synchrotron spectrum is shaped by electrons with Lorentz factor
$\gamma$ such that $\widehat\gamma_{\rm m}\,<\,\gamma\,<\,\gamma_0$,
which thus cool by interacting with photons in the range
$\nu\,<\,\nu_{\rm m}$; one can therefore use Eq.~(\ref{eq:bpslow})
with $1-\beta=4/3$, so that
\begin{eqnarray}
\quad 1-\beta'&\,\simeq\,&\frac{5-3p/2+\alpha_t/2}{3+\alpha_t/4}\quad
\left(\widehat\nu_{\rm m}\,<\,\nu\,<\,\nu_0\right)
\end{eqnarray}
The limit $\alpha_t\,\rightarrow\,0$ gives $\beta'=-2/3+p/2$, which
fits the results of Nakar et al. (2009) in this frequency range.

In the range ${\rm max}\left(\nu_{\rm c},\widehat\nu_{\rm
  c}\right)\,<\, \nu\,<\,{\rm min}\left(\widehat\nu_{\rm
  m},\nu_0\right)$, the Lorentz factor of electrons shaping that part
of the spectrum satisfies ${\rm max}\left(\gamma_{\rm
  c},\widehat\gamma_{\rm c}\right)\,<\, \gamma\,<\,{\rm
  min}\left(\widehat\gamma_{\rm m},\gamma_0\right)$, hence one can use
Eq.~(\ref{eq:bpslow}) with $1-\beta\,=\,(3-p)/2$, leading to
\begin{eqnarray}
1-\beta'&\,\simeq\,&\frac{7-3p+(1-p)\alpha_t}{4+(1-p)\alpha_t/2}
\nonumber\\
&& \left[{\rm max}\left(\nu_{\rm c},\widehat\nu_{\rm
    c}\right)\,<\, \nu\,<\,{\rm min}\left(\widehat\nu_{\rm
  m},\nu_0\right)\right]
\label{eq:bphnuc}
\end{eqnarray}
Note that $\widehat\nu_{\rm c}\,<\,\nu_{\rm c}$ is equivalent to
$\tilde \nu_{\rm c}\,<\,\nu_{\rm c}$. Here as well, one recovers the
index $\beta'\,=\,-3/4+3p/4$ derived in Nakar et al. (2009) in the
limit $\alpha_t\,\rightarrow\,0$.

If $\tilde \nu_{\rm c}\,<\,\nu_{\rm c}$, meaning $\widehat\gamma_{\rm
  c}\,<\,\gamma_{\rm c}$, the above completes the description of the
spectrum. If, however, $\nu_{\rm c}\,<\,\tilde \nu_{\rm c}$, one needs
to describe the intermediate range $\nu_{\rm
  c}\,<\nu\,<\,\widehat\nu_{\rm c}$. This range can actually be
decomposed into two sub-ranges, as follows.

For ${\widehat{\widehat\nu}}_{\rm c}\,<\,\nu\,<\,\widehat\nu_{\rm c}$,
the corresponding Lorentz factor verifies
${\widehat{\widehat\gamma}}_{\rm c}\,<\,\gamma\,<\,\widehat\gamma_{\rm
  c}$, hence $\nu_{\rm c}\,<\,\tilde\nu\,<\,\widehat\nu_{\rm c}$.
Consequently, the particle cools on the spectral range of $\nu
F_{\nu,\rm syn}$ that it contributes to through synchrotron emission,
so that one can use Eq.~(\ref{eq:bpslow}) with $\beta=\beta'$, giving
\begin{eqnarray}
1-\beta'&\,\simeq\,&\frac{-2+3\alpha_t+\left[4+(4-8p)\alpha_t+\alpha_t^2\right]^{1/2}}{2\alpha_t}
\nonumber\\
&&\quad\quad\quad \left(\widehat{\widehat\nu}_{\rm c}\,<\,\nu\,<\,\nu_{\rm
    c}\right)\label{eq:slowhathatnuc}
\end{eqnarray}
and $1-\beta'\,\simeq\,0.12$ for $\alpha_t=-0.5$, $p=2.3$, i.e. a
slowly rising $\nu F_{\nu,\rm syn}$.

Finally, in the remaining range $\nu_{\rm
  c}\,<\,\nu\,<\,{\widehat{\widehat\nu}}_{\rm c}$, the particle cools
on photons of frequency in the range $\widehat\nu_{\rm
  c}\,<\,\tilde\nu\,<\,\tilde\nu_{\rm c}$, for which $1-\beta$ is
given by Eq.~(\ref{eq:bphnuc}) above. This leads to
\begin{equation}
1-\beta'\,\simeq\,\frac{15-7p+\alpha_t(1-p)\left(10-p+\alpha_t\right)/2}
{8+\alpha_t(1-p)(10+\alpha_t)/4}
\label{eq:slowhathatnuc2}
\end{equation}
One finds $1-\beta'\,\simeq\,0.09$ for $\alpha_t=-0.5$ and $p=2.3$.

\subsubsection{Inverse Compton component}
In principle, one can derive analytical approximations to the
  inverse Compton component using the above broken power-law
  approximations, folded over the particle distribution, as in
  Eq.~(\ref{eq:fIC}) above. However, this appears rather intricate,
  given the number of potential power-law segments, in regards of the
  quality of the approximation that one can obtain; indeed, as
  discussed in detail in Sari \& Esin (2001), folding over the
  distribution function generally introduces logarithmic departures,
  which smooth out the power-law segments and breaks. One can
  nevertheless describe the general features of this inverse Compton
  component in the slow cooling regime as follows.

Below $\nu_{\rm IC,m}\,=\,2\gamma_{\rm m}^2\nu_{\rm m}$, the slope is
that of $\nu F_{\nu,\rm syn}$ below $\nu_{\rm m}$, i.e. 4/3. For
$\nu_{\rm IC,m}\,<\,\nu\,<\,\nu_{\rm IC,c}$ with $\nu_{\rm
  IC,c}\,=\,2\gamma_{\rm c}^2\nu_{\rm c}$, the slope of $\nu
F_{\nu,\rm IC}$ is $(3-p)/2$, reflecting that of $\nu F_{\nu,\rm syn}$
in the corresponding range. Above $\nu_{\rm IC,c}$, the inverse
Compton component reflects, up to the afore-mentioned logarithmic
corrections, the slope of $\nu F_{\nu,\rm syn}$ above $\nu_{\rm c}$,
which is generally close to flat or gently rising, see above.

Consequently, if the synchrotron spectrum has an extended range above
$\nu_{\rm c}$ where it is close to flat (i.e. $\beta\,\sim\,1$), then
one might have a close to flat inverse Compton component, at least up
to the cut-off frequency defined as
\begin{equation}
\nu_{\rm IC,KN}\,=\, \gamma_{\rm c}^2\tilde\nu_{\rm c}
\end{equation}
corresponding the boosting of $\tilde\nu_{\rm c}$ photons at the onset
of the Klein-Nishina regime by $\gamma_{\rm c}$ electrons. One can
define another cut-off frequency, as follows:
\begin{equation}
\nu_{\rm IC,\widehat c}\,=\,\widehat\gamma_{\rm c}^2\nu_{\rm c}
\end{equation}
which corresponds to the boosting of $\nu_{\rm c}$ photons by
electrons of Lorentz factor $\widehat\gamma_{\rm c}$, at the onset of
the Klein-Nishina regime. The ratio $\nu_{\rm IC,\widehat c}/\nu_{\rm
  IC,KN}$ can be written as $\widehat\gamma_{\rm c}/\gamma_{\rm c}$,
hence the ordering of one with respect to the other depends on whether
$\gamma\,<\,\widehat\gamma_{\rm c}$ or not. In any case, the actual
cut-off occurs at $\nu_{\rm IC,KN}$, while the presence of $\nu_{\rm
  IC,\widehat c}$ may lead to a feature (e.g. softening) in the
inverse Compton spectrum.  This can be understood by noting that in
the present case, the synchrotron (energy) flux generically peaks
above $\nu_{\rm c}$, while the particle distribution function falls
steeply beyond $\gamma_{\rm c}$, hence the peak of the inverse Compton
component is determined by the boosting of $\tilde \nu_c$ photons by
electrons of Lorentz factor $\gamma_{\rm c}$.

\subsubsection{Comparison to numerical calculations}
\begin{figure*}
\includegraphics[bb=250 0 1000 600, width=0.7\textwidth]{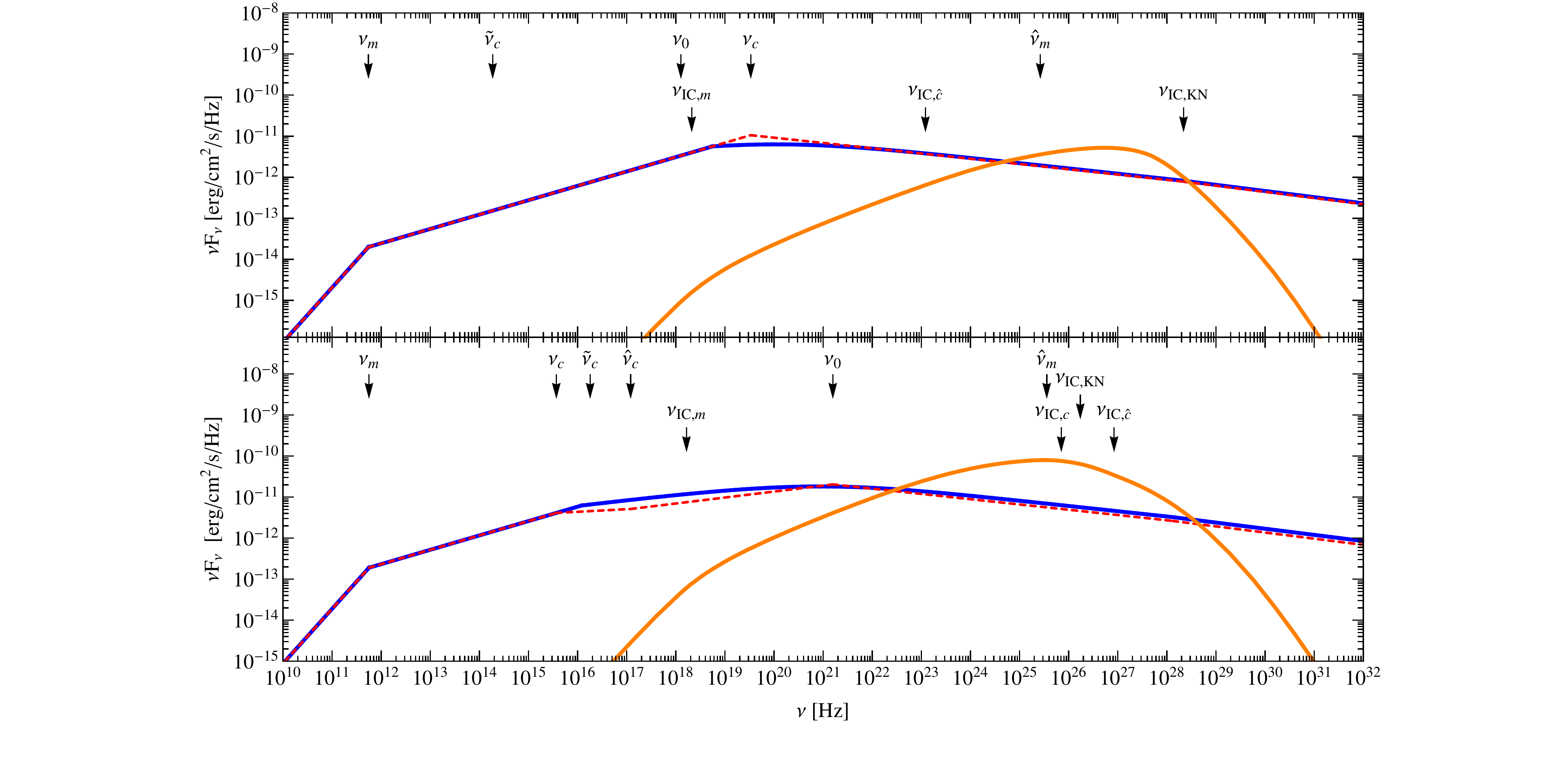}
\caption{Comparison of the analytical calculation (dashed red
  line) of the synchrotron spectrum in the slow cooling regime to a
  numerical calculation (solid blue line), for two representative
  cases: upper panel, observer time $t_{\rm obs}\,=\,10^4\,$s, blast
  energy $E=10^{53}\,$ergs, external density $n\,=\,0.01\,$cm$^{-3}$;
  lower panel, $t_{\rm obs}\,=\,3\times 10^4\,$s,
  $E\,=\,10^{54}\,$ergs, $n\,=\,10^{35}\,r^{-2}\,$cm$^{-3}$; in both
  cases, $\epsilon_e\,=\,0.1$, $p\,=\,2.3$ and
  $\epsilon_{B}\,=\,\epsilon_{B+}\left[t/(100\omega_{\rm
      pi}^{-1})\right]^{-0.4}$. The analytical estimates of the
  characteristic frequencies are indicated with arrows. The solid
  orange line represents the numerical calculation of the inverse
  Compton component. \label{fig:f1-slcomp} }
\end{figure*}

The above analytical broken power-law model of the synchrotron
  spectrum is compared to a full numerical calculation (with the
  algorithm described in Sec.~\ref{sec:fast} thereafter) in
  Fig.~\ref{fig:f1-slcomp}, in two different representative cases:
  upper panel, observer time $t_{\rm obs}\,=\,10^4\,$s, blast energy
  $E=10^{53}\,$ergs, external density $n\,=\,0.01\,$cm$^{-3}$; lower
  panel, $t_{\rm obs}\,=\,3\times 10^4\,$s, $E\,=\,10^{54}\,$ergs,
  $n\,=\,10^{35}\,r^{-2}\,$cm$^{-3}$ (wind profile with shock radius
  $r$ expressed in cm); for both, $\epsilon_e\,=\,0.1$, $p\,=\,2.3$,
  $\epsilon_{B+}\,=\,0.01$ and $\alpha_t\,=\,-0.4$, assuming
  $\epsilon_{B}\,=\,\epsilon_{B+}\left[t/(100\omega_{\rm
      pi}^{-1})\right]^{\alpha_t}$. For a decelerating adiabatic
  Blandford \& McKee (1976) solution, the value of the blast Lorentz
  factor at these observer times are $\Gamma_{\rm b}\,\simeq\,33$ for
  the upper panel and $\Gamma_{\rm b}\,\simeq\,29$ in the lower
  panel. For the above decay law of the magnetic field, one finds for
  the first scenario $\epsilon_{B-}\,=\,3.2\times 10^{-5}$ ($t_{\rm
    dyn}\,=\,1.3\times 10^6\,$s, $\omega_{\rm pi}^{-1}\,=\,7.6\times
  10^{-3}\,$s), and in the second scenario
  $\epsilon_{B-}\,=\,2.1\times 10^{-5}$ ($t_{\rm dyn}\,=\,1.7\times
  10^6\,$s, $\omega_{\rm pi}^{-1}\,=\,3.5\times 10^{-3}\,$s).

The critical frequencies are indicated with arrows. The thick solid
line corresponds to the numerical calculation (synchrotron in blue,
inverse Compton component in orange) while the dashed line shows the
analytical estimates, which clearly provides a faithful match in both
cases.

In the first scenario (upper panel), $\widehat\gamma_{\rm
  c}\,\,\simeq\,60$ while $\gamma_{\rm c}\,\simeq\,10^7$:
Klein-Nishina effects are therefore particularly strong; there are
actually so strong that $\nu_0\,<\,\nu_{\rm c}$, which means that the
Klein-Nishina suppression of electron cooling reduces the Compton
parameter to below unity at $\gamma_{\rm c}$. Consequently, the
dependence of $Y$ on $\gamma$ does not affect the spectrum above
$\nu_{\rm c}$. Noting that in this region $\nu F_\nu\,\propto
\gamma^{2-p}$ [Eq.~(\ref{eq:nuFnuhi})] and
$\nu\,\propto\,\gamma^{2-\alpha_t/2(1+\alpha_t)}$
      [Eq.~(\ref{eq:nuscal})], one finds $\nu
      F_\nu\,\propto\,\nu^{1-\beta}$ with
      $1-\beta\,=\,(2-p)/[2-\alpha_t/2(1+\alpha_t)]$, which matches
      Eq.~(\ref{eq:slowsupnu0}). As mentioned earlier, the spectrum
      remains unaffected with respect to the standard synchrotron
      spectrum below $\nu_{\rm c}$ because the electrons shaping that
      part of the spectrum do not cool on a dynamical timescale.

In this first scenario, the method proposed in
Eqs.~(\ref{eq:Ycgcslow}) overestimates $\gamma_{\rm c}$ by a factor
$2.5$, hence $\nu_{\rm c}$ by a factor $6$ and $\nu_{\rm
  IC,KN}\,=\,\tilde\nu_{\rm c}\gamma_{\rm c}^2$ by a factor $2.5$ as
well. Taking into account this overestimate, the numerical calculation
indicates that the suppression of the inverse Compton flux becomes
noticeable at a factor $\sim\,5$ below the theoretical value of
$\nu_{\rm IC,KN}$, as calculated with the correct $\gamma_{\rm c}$. As
anticipated, the characteristic frequency $\nu_{\rm IC,\widehat c}$
leads to a soft softening in the inverse Compton component, but not to
a cut-off.

In the second scenario (lower panel), $\gamma_{\rm
  c}\,<\,\widehat\gamma_{\rm c}$ hence Klein-Nishina suppression of the
inverse Compton cooling becomes effective at $\widehat\nu_{\rm c}$
only. The (analytical) synchrotron spectrum is thus close to flat in
the region $\nu_{\rm c}\,\lesssim\,\nu\,\lesssim\,\widehat\nu_{\rm c}$, as
indicated by Eqs.~(\ref{eq:slowhathatnuc2}), (\ref{eq:slowhathatnuc})
above, then rising with $1-\beta\,\simeq\,0.14$ corresponding to
Eq.~(\ref{eq:bphnuc}) above for the range $\widehat \nu_{\rm
  c}\,<\,\nu\,<\,\nu_0$, because $\nu_0\,<\,\nu_{\rm m}$. Above
$\nu_0$, Eq.~(\ref{eq:slowsupnu0}) applies and gives the same high
energy spectral slope as for the previous scenario.

In this second case, the analytical calculations underestimate
$\gamma_{\rm c}$ by a factor $1.8$. There is nevertheless a broad
satisfactory agreement between the analytical synchrotron spectrum and
the numerical calculation.

\subsubsection{Comparison to non-decaying scenarios}
\begin{figure*}
\includegraphics[bb=130 0 430 300, width=0.4\textwidth]{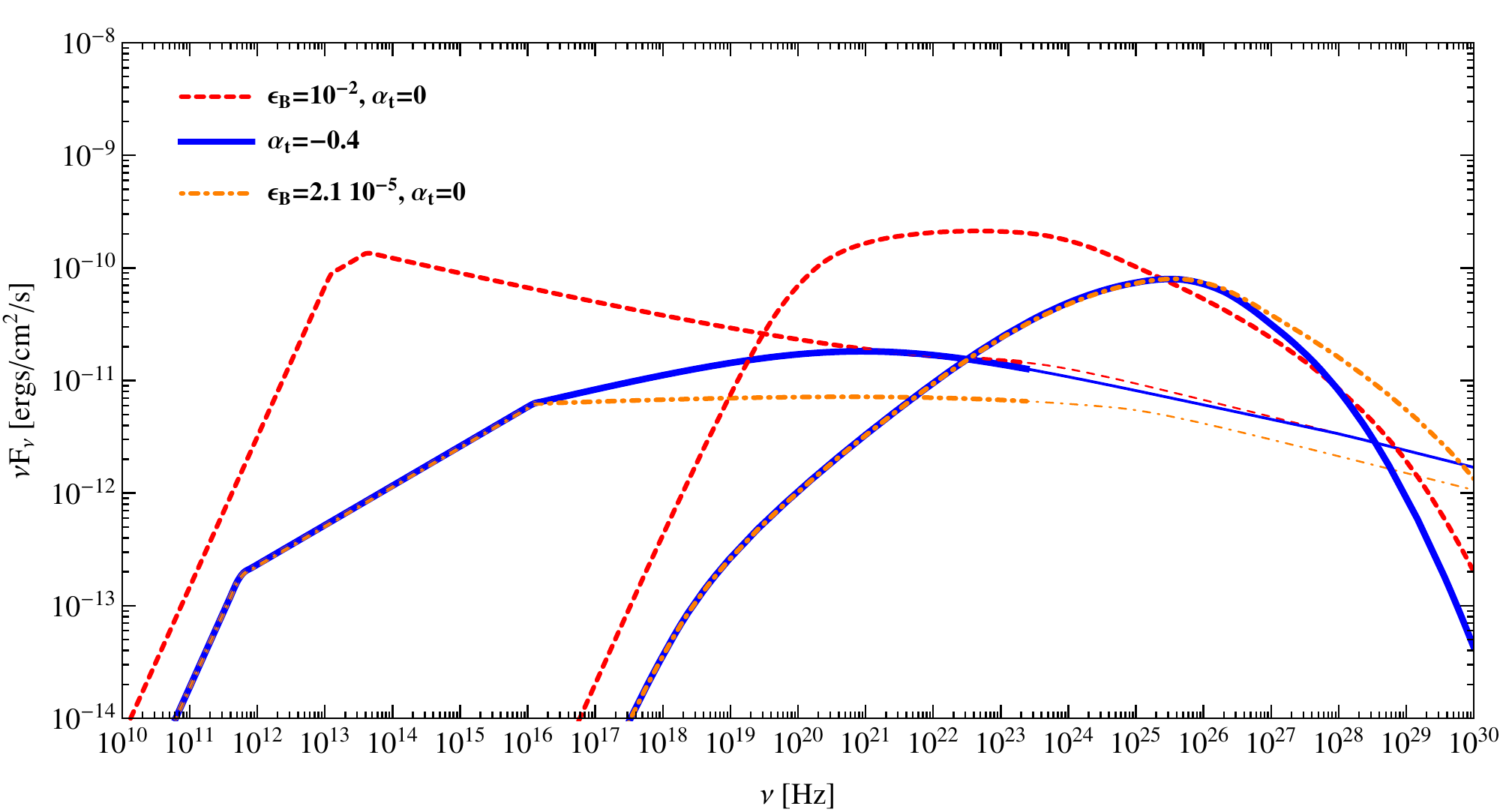}
\caption{Synchrotron and inverse Compton spectra at $t_{\rm
    obs}=3\times10^4\,$s, representative of the slow cooling regime,
  for a blast with energy $E=10^{54}\,$ergs impinging on a progenitor
  wind with density $n=10^{35}\,r^{-2}\,$cm$^{-3}$ ($r$ in cm),
  assuming $\epsilon_e=0.1$ and $p=2.3$, for three microphysical
  models, as indicated: homogeneous (non-decaying)
  $\epsilon_B\,=\,\epsilon_{B+}\,=\,0.01$ (dashed red line); decaying
  $\epsilon_B\,=\,\epsilon_{B+}\left[t/(100\omega_{\rm
      pi}^{-1})\right]^{-0.4}$ (solid blue); homogeneous
  $\epsilon_B\,=\,\epsilon_{B-}\,=\,2.1\times10^{-5}$ (dash-dotted
  orange), the value of $\epsilon_{B-}$ being representative of
  $\epsilon_B$ close to the contact discontinuity in the decaying
  $\epsilon_B$ model. The synchrotron model predictions have been
  thinned beyond an ad-hoc maximal synchrotron photon energy of
  $1\,{\rm GeV}$ (see text). Characteristic frequencies are: for
  $\alpha_t=-0.4$ and for $\epsilon_B\,=\,\epsilon_{B-}$, $\nu_{\rm
    m}\,\simeq\,5.6\times 10^{11}\,$Hz, $\nu_{\rm
    c}\,\simeq\,1.2\times 10^{16}\,$Hz and $\nu_{\rm
    IC,KN}\,\simeq\,2\times 10^{26}\,$Hz; for
  $\epsilon_{B}\,=\,\epsilon_{B+}$, $\nu_{\rm m}\,\simeq\,1.2\times
  10^{13}\,$Hz, $\nu_{\rm c}\,\simeq\,4.2\times 10^{13}\,$Hz and
  $\nu_{\rm IC,KN}\,\simeq\,2\times 10^{24}\,$Hz.
  \label{fig:f1-slow} }
\end{figure*}

Figure~\ref{fig:f1-slow} provides a numerical comparison of the
spectra shown in the lower panel of Fig.~\ref{fig:f1-slcomp} with two
calculations for the same parameters but a homogeneous (non-decaying)
turbulence: one in which $\epsilon_B\,=\,\epsilon_{B+}\,=\,0.01$,
another one in which $\epsilon_B\,=\,\epsilon_{B-}\,=\,2.1 \,\times
10^{-5}$, which corresponds to the value of $\epsilon_{B}(t_{\rm
  dyn})$, i.e. close to the contact discontinuity, in the above
decaying micro-turbulence model.

 As expected the SSC spectrum with decaying microturbulence merges
 with that corresponding to uniform $\epsilon_{B-}$ at frequencies
 below $\nu_{\rm c}$, since electrons of Lorentz factor
 $\gamma<\gamma_{\rm c}$ then cool in magnetized turbulences of equal
 strength in both models. The synchrotron spectrum for decaying
 micro-turbulence also merges with the synchrotron spectrum for
 uniform $\epsilon_{B+}$ at the highest frequencies, since the cooling
 time for those emitting electrons becomes shorter than $\Delta$,
 hence the particles effectively cool in a magnetic field
 characterized by $\epsilon_{B+}$. However, the integrated powers for
 these two models differ, because the cooling efficiencies
 $\sim\,\left(\gamma_{\rm c}/\gamma_{\rm m}\right)^{2-p}$ differ.

In this regard, the slow cooling synchrotron spectrum for decaying
micro-turbulence is a hybrid of the spectra for uniform high and low
$\epsilon_B$, transiting from $\epsilon_{B-}$ at low frequencies to
$\epsilon_{B+}$ at high frequencies. This justifies the use of a two
zone model, one with low $\epsilon_{B-}$ and one for high
$\epsilon_{B+}$, to compute an approximated spectrum in wavebands at
respectively low and high frequencies.
 
In Fig.~\ref{fig:f1-slow}, the synchrotron spectra have been
arbitrarily continued at very high frequencies, albeit with a thin
line, but they should of course cut-off at some maximal energy where
the acceleration timescale becomes of the same order as the cooling
timescale. Since this depends on acceleration physics, see the
discussion in Lemoine (2013), the lines have been turned from thick to
thin at an ad-hoc location corresponding to a synchrotron photon
energy of $1\,$GeV. This estimate is discussed in Plotnikov et
al. (2013), Lemoine (2013), Wang et al. (2013), Sironi et al. (2013);
it depends on the afterglow parameters and, in particular, on observer
time.

For reference, one notes the critical frequencies:
\begin{eqnarray}
\nu_{\rm IC,c}&\,=\,&2\gamma_{\rm c}^2\nu_{\rm
  c}\nonumber\\ &\,\simeq\,& 7.3\times 10^{23}\,{\rm
  Hz}\,E_{54}\epsilon_{B-,-5}^{-7/2}A_{\star,11.7}^{-9/2}t_{\rm
  obs,4.5}^2z_+^{-3}Y_{\rm c,2}^{-4} \nonumber\\ \nu_{\rm
  IC,KN}&\,=\,& \frac{1}{1+z}\Gamma_{\rm b}\gamma_{\rm c}m_e
c^2\nonumber\\ &\,\simeq\,& 4\times 10^{25}\,{\rm Hz}\,
E_{54}^{1/2}\epsilon_{B-,-5}^{-1}A_{\star,11.7}^{-3/2}t_{\rm
  obs,4.5}^{1/2}z_{+}^{-3/2}Y_{\rm
  c,2}^{-1}\nonumber\\ &&\label{eq:ICnu}
\end{eqnarray}
with the notations $z_{+}\,=\,(1+z)/2$, $Y_{\rm c,2}\,=\,(1+Y_{\rm
  c})/100$ and $A_{\star,11.7}\,=\,A_{\star}/(5\times10^{11}\,{\rm
  g/cm^2})$. For reference, in the scenario of Figs.~\ref{fig:f1-slow}
and \ref{fig:f1-fast}, $A_{\star}\,\simeq\,0.3$ and $Y_{\rm
  c}\,\simeq\,50$ for $\epsilon_{B-}=2.1\times 10^{-5}$.  Note the
strong dependence of $\nu_{\rm IC,c}$ and $\nu_{\rm KN}$ on the
external density. As discussed above, the peak of the inverse Compton
component is expected to occur at $\nu_{\rm IC,KN}$, although the
numerical calculation suggests that the turn-over becomes manifest a
factor $\sim 5$ below the above theoretical value.

The VERITAS collaboration has recently been able to observe the
exceptional GRB130427A and to put stringent upper limits on the
emission at $\gtrsim100\,$GeV (Aliu et al. 2014). The absence of
detection of this energy range suggests that, for this burst at least,
the inverse Compton component has cut-off below $\simeq100\,$GeV,
while the Fermi detection of multi-GeV photons up to a day or so
suggests that this cut-off lied above $1-10\,$GeV. Such a cut-off
energy fits well with the above estimates for $\nu_{\rm IC,KN}$ for a
low average $\epsilon_B$.

\subsection{Fast cooling}\label{sec:fast}

The fast cooling regime involves a susbtantial variety of synchrotron
spectra, with multiple breaks and indices, as discussed in detail in
Nakar et al. (2009) and Wang et al. (2010) for the case of uniform
$\epsilon_B$. One key difference with the slow-cooling regime is that
particles with Lorentz factors $\gamma\,<\,{\rm max}\left(\gamma_{\rm
  c},\gamma_{\rm m}\right)$ may have a non-trivial cooling history
while in the slow-cooling regime, such particles do not cool.  As a
consequence, it is difficult to even derive the cooling Lorentz factor
and the Compton parameter $Y_{\rm c}$ when Klein-Nishina effects
become significant.

In this fast-cooling regime, it is actually more efficient to compute
the spectrum numerically, using the following simple and efficient
algorithm. One starts with a template synchrotron spectrum, for
instance that corresponding to a homogeneous magnetized
turbulence. One can then derive a first approximation to $\gamma_{\rm
  c}$ and $Y_{\rm c}$, either using standard formulae (e.g Panaitescu
\& Kumar 2000) -- which ignore KN effects -- or through an explicit
determination of $\gamma_{\rm c}$ as the Lorentz factor for which
cooling takes place on a dynamical timescale, using a radiation energy
density inferred from the template spectrum. With $\gamma_{\rm c}$ and
$Y_{\rm c}$, one can solve Eqs.~(\ref{eq:UBscal}), (\ref{eq:nuscal})
and (\ref{eq:Yscal}) to compute the frequencies and Compton parameter
as a function of the initial Lorentz factor of an electron; one can
then use Eqs.~(\ref{eq:nuFnuhi}) and (\ref{eq:nuFnulo}) to compute an
improved version of the synchrotron spectrum, properly taking into
account the cooling history of the electrons in the decaying
turbulence as well as all relevant Klein-Nishina effects. This latter
spectrum remains an approximation, because it relies on a guessed
value for $\gamma_{\rm c}$ and $Y_{\rm c}$. Nevertheless, iterating
the above process, using each time as a template the previously
computed synchrotron spectrum, one obtains after $\,\sim\,10$
iterations a self-consistent synchrotron spectrum, with $\gamma_{\rm
  c}$ and $Y_{\rm c}$ determined to high accuracy.  Finally, one can
derive the inverse Compton spectrum using Eq.~(\ref{eq:fIC}).

\begin{figure*}
\includegraphics[bb=130 0 430 300, width=0.4\textwidth]{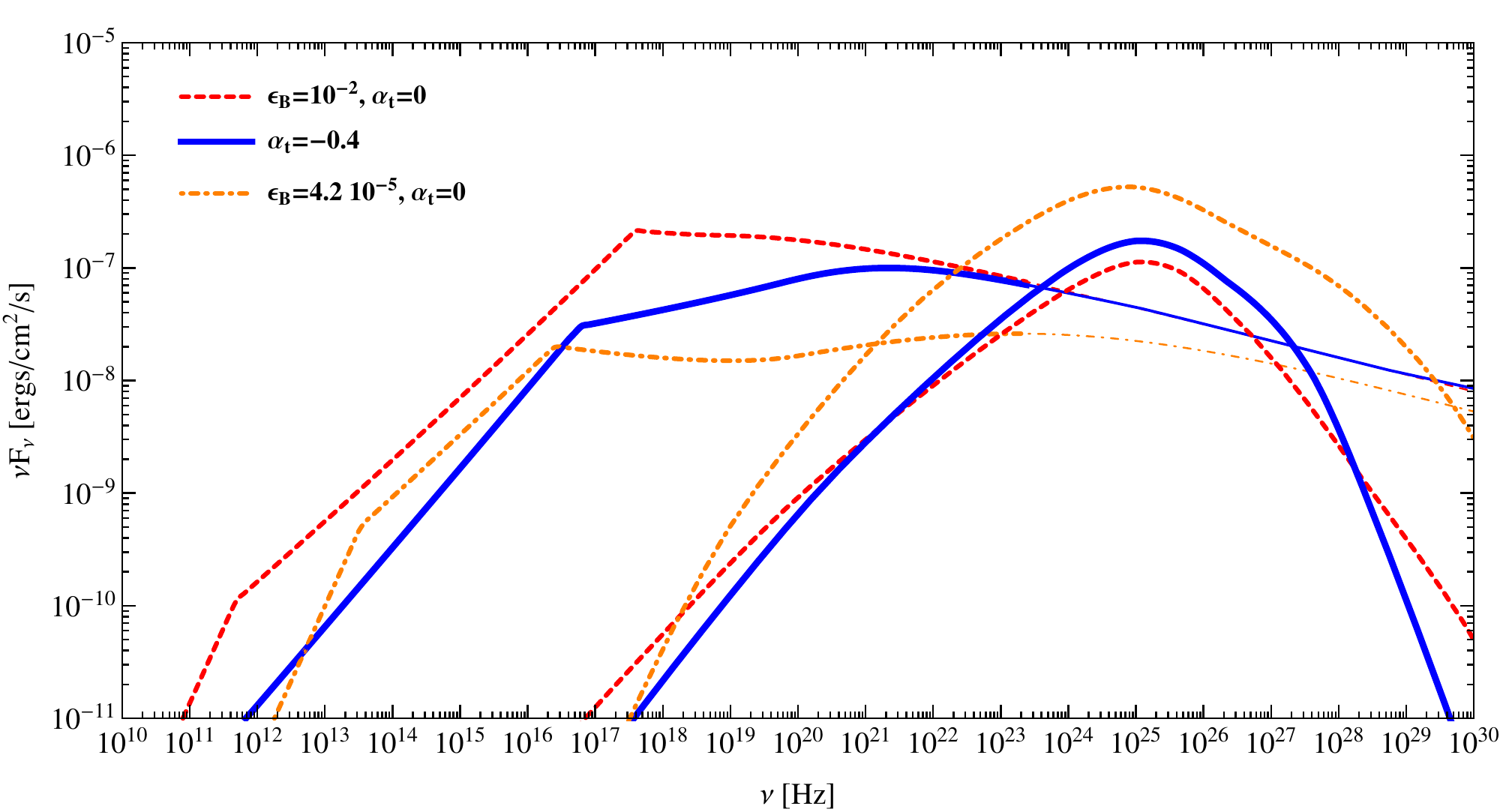}
\caption{Same as Fig.~\ref{fig:f1-slow} at observer time $t_{\rm
    obs}=30\,$s, representative of the fast cooling regime.
  Characteristic frequencies are: for $\alpha_t=-0.4$, $\nu_{\rm
    c}\,\simeq\,10^{12}\,$Hz, $\nu_{\rm m}\,\simeq\,3\times
  10^{16}\,$Hz and $\gamma_{\rm m}^2\tilde\nu_{\rm
    m}\,\simeq\,6\times10^{25}\,$Hz; for
  $\epsilon_B\,=\,\epsilon_{B-}$, $\nu_{\rm
    c}\,\simeq\,2\times10^{13}\,$Hz, $\nu_{\rm m}\,\simeq\,3\times
  10^{16}\,$Hz and $\gamma_{\rm m}^2\tilde\nu_{\rm
    m}\,\simeq\,6\times10^{25} \,$Hz; for
  $\epsilon_{B}\,=\,\epsilon_{B+}$, $\nu_{\rm c}\,\simeq\,4\times
  10^{11}\,$Hz, $\nu_{\rm m}\,\simeq\,4\times 10^{17}\,$Hz and
  $\gamma_{\rm m}^2\tilde\nu_{\rm m}\,\simeq\,6\times10^{25}\,$Hz.
  \label{fig:f1-fast} }
\end{figure*}

The above algorithm provides a self-consistent estimate of the
synchrotron and inverse Compton spectra with a normalization accuracy
of order unity. This accuracy can be checked by calculating a
posteriori the integrated synchrotron and inverse Compton powers and
comparing to the total electron power injected through the shock: in
the fast cooling regime, these should match. The error is of order
$10-40\,$\% for the SSC spectrum of a decaying micro-turbulence as
shown in Fig.~\ref{fig:f1-fast}, depending on observer time; it is
less than $10-20\,$\% for $\alpha_t\,=\,0$ and
$\epsilon_B\,=\,\epsilon_{B+}$, but it becomes a factor $\lesssim 2$
for $\alpha_t\,=\,0$ and $\epsilon_B\,=\,\epsilon_{B-}$. Most of the
error results from the broken power-law normalization of the flux in
Eq.~(\ref{eq:nuFnuhi}) and from the treatment of the Klein-Nishina
cross-section as a step function in the calculation of the synchrotron
cooling history. The resulting uncertainty remains nevertheless
satisfactory given the uncertainty associated for instance to the
definition of $t_{\rm dyn}$ (hence $\gamma_{\rm c}$) in the absence of
a realistic description of the blast energy profile.

A detailed example of the SSC spectrum of the blast, for the same
parameters as in Fig.~\ref{fig:f1-slow}, is presented in
Fig.~\ref{fig:f1-fast} at an observer time $t_{\rm obs}\,=30\,$s;
assuming a Blandford \& McKee (1976) decelerating solution in a wind
profile, the Lorentz factor of the blast at that time is $\Gamma_{\rm
  b}\,\simeq\,160$. The value of $\epsilon_{B-}$ in this case is
$4.2\times 10^{-5}$.  As expected, the spectrum corresponding to a
decaying turbulence merges with the spectrum for uniform
$\epsilon_B=\epsilon_{B+}$ above a frequency $\nu\,\sim\,10^{23}\,$Hz,
since the electrons that emit in that range cool fast, in a region
where $\epsilon_B\,\simeq\,\epsilon_{B+}$. In this fast cooling
regime, the total integrated energy densities of the three SSC spectra
correspond to the injected electron energy density. The spectrum for
decaying micro-turbulence does not merge with that for uniform
$\epsilon_B\,=\,\epsilon_{B-}$ at low frequencies, since the cooling
Lorentz factors differ for both. In this fast cooling regime, one
cannot therefore describe accurately the synchrotron spectrum at low
frequencies with a spectrum computed for uniform low $\epsilon_{B}$:
an explicit calculation becomes necessary.

Finally, note that the inverse Compton component in the fast cooling
regime is expected to peak at $\gamma_{\rm m}^2\nu_{\rm m}$, if the
synchrotron flux peaks at $\nu_{\rm m}$ and if one omits KN
effects. The Klein-Nishina suppression implies a turn-over of the IC
component at most at $\gamma_{\rm m}^2\tilde\nu_{\rm m}$, for reasons
analog to those discussed in the slow cooling regime, see also Nakar
et al. (2009). If $\nu_{\rm m}\,<\,\tilde\nu_{\rm m}$, the slow rise
of the synchrotron flux above $\nu_{\rm m}$ in the case of low
$\epsilon_B$ (due to the KN suppression of electron cooling) or
decaying microturbulence implies a comparable behavior of the IC
component between $\gamma_{\rm m}^2\nu_{\rm m}$ and $\gamma_{\rm
  m}^2\tilde\nu_{\rm m}$. This feature is not clearly seen in
Fig.~\ref{fig:f1-fast} due to the (relative) proximity of these two
frequencies, $\gamma_{\rm m}^2\tilde\nu_{\rm
  m}\,\simeq\,6\times10^{25}\,$Hz and $\gamma_{\rm m}^2\nu_{\rm
  m}\,\simeq\,10^{24}\,$Hz.

\section{Spectra and light curves}\label{sec:lc}

\subsection{Spectral and temporal behaviors}
The spectra shown in Figs.~\ref{fig:f1-slow} and \ref{fig:f1-fast}
illustrate how a complete and simultaneous spectral coverage would
allow to tomograph the evolution of the micro-turbulence behind the
relativistic shock. The effects are most noticeable in the X-ray and
MeV regions, as one would expect: in this region of the spectrum, the
emitting electrons feel the decaying turbulence, while at the highest
frequencies, they cool in regions with
$\epsilon\,\sim\,\epsilon_{B+}$, and at frequencies $\nu\,<\,\nu_{\rm
  c}$ they cool in a magnetic field characterized by $\epsilon_{B-}$,
the value of which evolves slowly in time.

\begin{figure}
\includegraphics[bb=30 120 500 1350, width=0.45\textwidth]{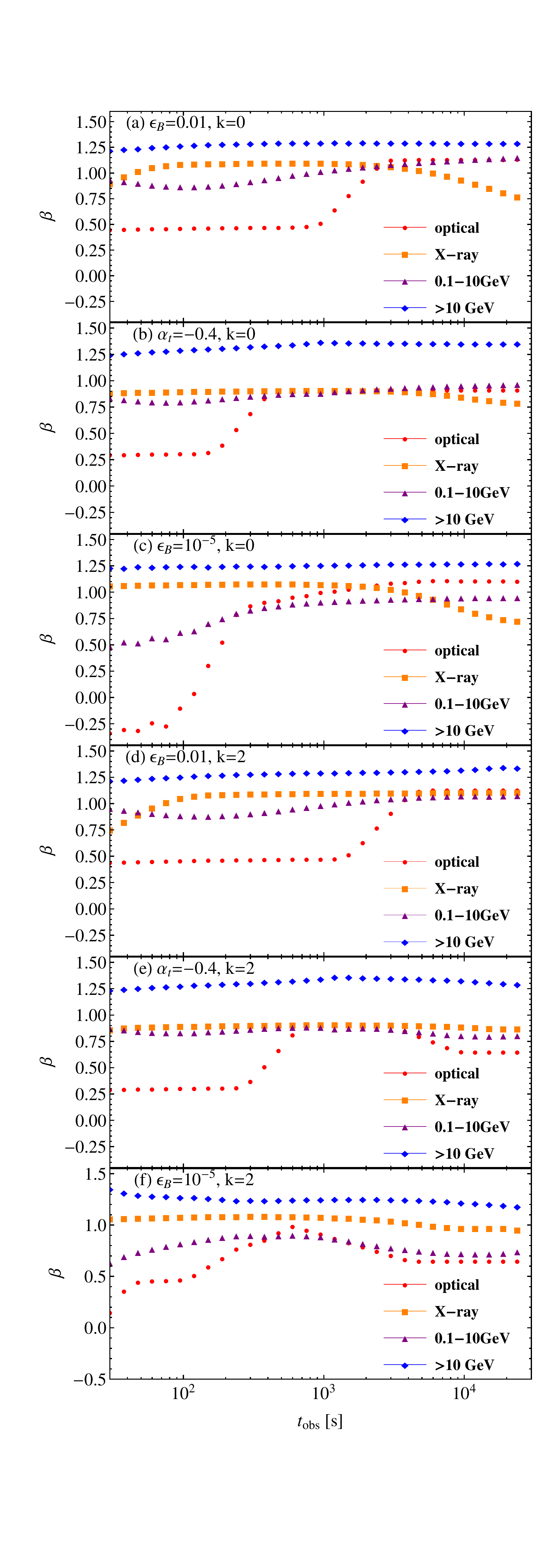}
\caption{Flux density index $\beta$, with
  $F_\nu\,\propto\,\nu^{-\beta}$, as a function of time, in various
  wavebands, for different external density profiles: $k=2$ (wind,
  $n=10^{35}r^{-2}\,$cm$^{-3}$) and $k=0$ (constant density,
  $n=1\,$cm$^{-3}$), assuming uniform $\epsilon_{B}=0.01$,
  $\epsilon_B\,=\,(100\omega_{\rm pi}t)^{\alpha_t}$ or uniform
  $\epsilon_{B}=10^{-5}$. Blast energy, jet Lorentz factor and
  $\epsilon_e$ are as in Fig.~\ref{fig:f1-slow}. Wavebands are as
  indicated: for reference, optical (circles) $\nu\,=\,4.7\times
  10^{14}\,$Hz, X-ray (squares) integrated from 0.3 to 10~keV
  (i.e. $0.72-24\times 10^{17}\,$Hz), 0.1-10~GeV (upward triangles, in
  frequency $0.24-24\times 10^{23}\,$Hz) and $>\,10\,$GeV (diamonds,
  in frequency $>\,24\times 10^{23}\,$Hz).
  \label{fig:f2-beta} }
\end{figure}

Afterglow models often rely on the spectral and temporal slopes in
various domains and their so-called closure relations to make
comparison to observations.  Figure~\ref{fig:f2-beta} therefore
presents the spectral slopes $\beta$ defined by
$F_{\nu}\,\propto\,\nu^{-\beta}$ (where $F_{\nu}$ sums the synchrotron
and inverse Compton fluxes), for the optical, X-ray, high energy
($0.1-10\,{\rm GeV})$ and very high energy ($>10\,{\rm GeV}$) ranges
(note that no attenuation in extra-galactic background radiation has
been assumed for the latter range). This figure compares the spectral
slopes for the three previous representative models:
($\alpha_t\,=\,0$, $\epsilon_B\,=\,0.01$), $\alpha_t\,=\,-0.4$ and
($\alpha_t\,=\,0$, $\epsilon_B\,=\,10^{-5}$) with otherwise same
parameters as in Figs.~\ref{fig:f1-slow}, \ref{fig:f1-fast}, except
$k$, which takes values $0$ (constant density profile) or $2$ (stellar
wind).

In the optical, $\beta$ has been calculated at a reference frequency
of $4.7\times10^{14}\,$Hz (R band); in the X-ray, $\beta$ is calculated
as the average slope over the interval of energies $0.3-10\,$keV; at
high energy, it is calculated as the average of the energy interval
$0.1-10\,$GeV and at very high energy, over $>10\,$GeV.

According to Fig.~\ref{fig:f2-beta}, the most robust signature of a
decaying micro-turbulence appears to be a slightly harder slope in the
X-ray, $\beta\,\simeq\,0.9$ [panels (b) and (e)] vs
$\beta\,\simeq\,1.15$ for uniform $\epsilon_B$ in the first hours
[panels (a), (c), (d) and (f)]. The latter value corresponds to the
fast cooling regime $\beta=p/2$, therefore it depends on $p$; however,
one does not expect it to go below $1$, because $p>2$ is a generic
prediction of relativistic shock acceleration, e.g. Bednarz \&
Ostrowski (1998), Kirk et al. (2000), Achterberg et al. (2001),
Lemoine \& Pelletier (2003), Sironi et al. (2013). Current data do not
allow to distinguish between these limits; in particular, the Swift
data lead to $\beta\,\simeq\,1\,\pm0.1$ (Evans et al. 2009) in
afterglows with standard power-law decay. Interestingly, even in the
case of a homogeneous turbulence, the X-ray slope hardens at late
times because of the emergence in the X-ray range of the inverse
Compton component; for instance, in panels (a) and (c), one can see
$\beta$ transit to values of order 0.7, corresponding to the low
energy extension of this inverse Compton component with slope
$\beta\,=\,(p-1)/2$.

In the optical range, the slope is comparable to that in the X-ray
range for the decaying micro-turbulence scenario at observer times
$\sim\,10^3\,$s, but significantly harder at earlier times when the
optical falls in the range $\nu_{\rm c}\,-\nu_{\rm m}$:
Fig.~\ref{fig:f2-beta} indicates values $\beta\,\sim\,0.3$, harder
than expected $(1/2)$ in the standard fast cooling regime in this
range of frequencies. At an observer time $t_{\rm obs}\,=30\,$s, for
$k=2$ corresponding to panel (e) as well as to Fig.~\ref{fig:f1-fast},
$\tilde\nu_{\rm m}\,\simeq\,1.4\times 10^{18}\,$Hz, which lies below
the peak of the synchrotron component (see
Fig.~\ref{fig:f1-fast}). This implies that Klein-Nishina effects are
significant, and that $\gamma_{\rm m}$ electrons cool by interacting
with the segment in the range $10^{17}\,-10^{21}\,$Hz, whose index
$\beta\,\sim\,0.8$. In this case, one can compute the expected index
$\beta'$ of the segment below $\nu_{\rm m}$, using a variant of
Eq.~(\ref{eq:bpslow}): one notes that $\nu F_{\nu,\rm syn}\,\propto\,
\gamma/(1+Y)$ [Eq.~(\ref{eq:nuFnulo})], which gives
\begin{equation}
1-\beta'\,=\,\frac{1-\alpha_t+(1-\beta)(1+\alpha_t)}{2-\alpha_t/2 +
  (1-\beta)\alpha_t/2}\,\simeq\,0.3
\end{equation}
the last equality applying for $\alpha_t\,=\,-0.4$ and
$\beta\,=\,0.8$. This explains the values of $\alpha_t$ found in the
optical range. Such values do depart from the standard synchrotron
spectra, although Klein-Nishina effects may also cause values
$\beta\,\sim\,0.3$ below $\nu_{\rm min}$ in the case of homogeneous
turbulence scenarios, see Nakar et al. (2009), their figures~2 and 3
for example. Therefore, it is not clear at present whether one can
consider such values of $\beta$ as a clear signature of a decaying
micro-turbulence.

All in all, an accurate measurement of $\beta$ in the X-ray range, or
better, in the MeV range if the MeV afterglow could be detected, would
provide the best probe of $\alpha_t$, see also Fig.~\ref{fig:f1-slow}
and \ref{fig:f1-fast} for illustrations of these effects.

Another quantity of interest for a general description of the
afterglow is the temporal slope, defined by $F_{\nu}\,\propto\,t_{\rm
  obs}^{-\alpha}$. Of course, these temporal slopes directly depends
on the time evolution of the various parameters, through the assumed
evolutionary law for $\Gamma_{\rm b}$ and the evolution of $r$ and
$n$, in contrary to the spectral slopes $\beta$ at any given
time. Here $\Gamma_{\rm b}$ is assumed to decrease as in the Blandford
\& McKee (1976) adiabatic solution, i.e. $\Gamma_{\rm b}\,\propto\,
t^{(k-3)/[2(4-k)]}$.

\begin{figure}
\includegraphics[bb=30 120 500 1350, width=0.45\textwidth]{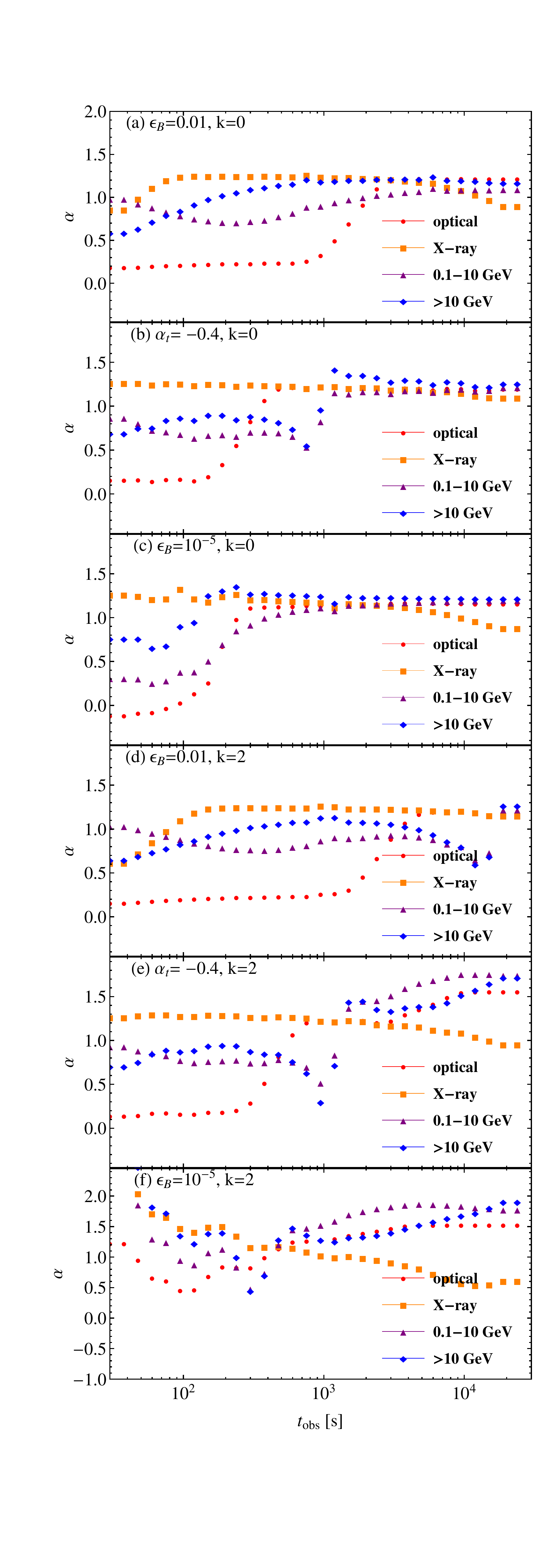}
\caption{Observer time decay index of the flux density,
  i.e. $F_\nu\,\propto\,t_{\rm obs}^{-\alpha}$, for various models, as
  indicated in Fig.~\ref{fig:f2-beta}. Blast energy and $\epsilon_e$
  are as in Fig.~\ref{fig:f1-slow}.
  \label{fig:f3-alpha} }
\end{figure}

The values of $\alpha$ for the same models and intervals as
Fig.~\ref{fig:f2-beta} are reported in Fig.~\ref{fig:f3-alpha}. The
largest differences between constant (high) and decaying $\epsilon_B$
result from the different transit times between the slow and fast
cooling regimes, but these times depend in turn on other parameters
that are a priori unknown. The light curves otherwise present similar
features, without a clear trend distinguishing one from the other.

Figures~\ref{fig:f2-beta} and \ref{fig:f3-alpha} thus indicate that
$\alpha$ and $\beta$ are by themselves weakly sensitive probes of the
dynamics of the magnetized turbulence in the blast and that it is not
possible at present to distinguish a decaying micro-turbulence from a
uniform low- or high $\epsilon_B$ on the basis of these data. It
appears much more effective to try to probe $\epsilon_B$ through a
multiwavelength fit of the afterglow light curves, using not only the
temporal and spectral slopes, but also the ratio of fluxes between
different spectral windows, as done in Lemoine et al. (2013), Liu et
al. (2013) for Fermi-LAT bursts.

\subsection{Emission at very high energy}
Finally, an interesting consequence of a decaying micro-turbulence is
the generic prediction of substantial emission at the highest
energies, due to the large value of the Compton parameter. Naive
estimates, $Y\,\sim\,\sqrt{\epsilon_e/\epsilon_{B-}}$, indicate values
of several hundreds for $Y$, although they neglect Klein-Nishina
effects which depend on the electron energy, therefore on observed
frequency; furthermore, the non-trivial spectral shape above $\nu_{\rm
  c}$ in the case of decaying micro-turbulence modifies the ratio of
the inverse Compton to the synchrotron component. As discussed in Wang
et al. (2013), emission above $10\,$GeV is most likely of inverse
Compton origin, because the maximal synchrotron photon energy is more
likely of the order of $1\,$GeV or so at $100-1000\,$sec observer
time. This emission is a prime target for future gamma-ray telescopes
such as HAWK (Mostafa 2013) or CTA (Inoue et al. 2013). For the
particular case of CTA, Inoue et al. (2013) have investigated the
detection rates of gamma-ray bursts above $30\,$GeV by assuming that
the spectrum continues beyond $1\,$GeV with a spectral index
$\beta=1.1$ (corresponding to the standard fast cooling regime
$\beta=p/2$ with $p=2.2$) and scaling the flux at $1\,$GeV to that
measured by the Fermi-LAT instruments. Their simulations lead to about
one detection per year. This rate is rather low, therefore any
improvement would be quite valuable, given the potential impact of a
high energy detection.

\begin{figure}
\includegraphics[bb=-20 0 420 280, width=0.49\textwidth]{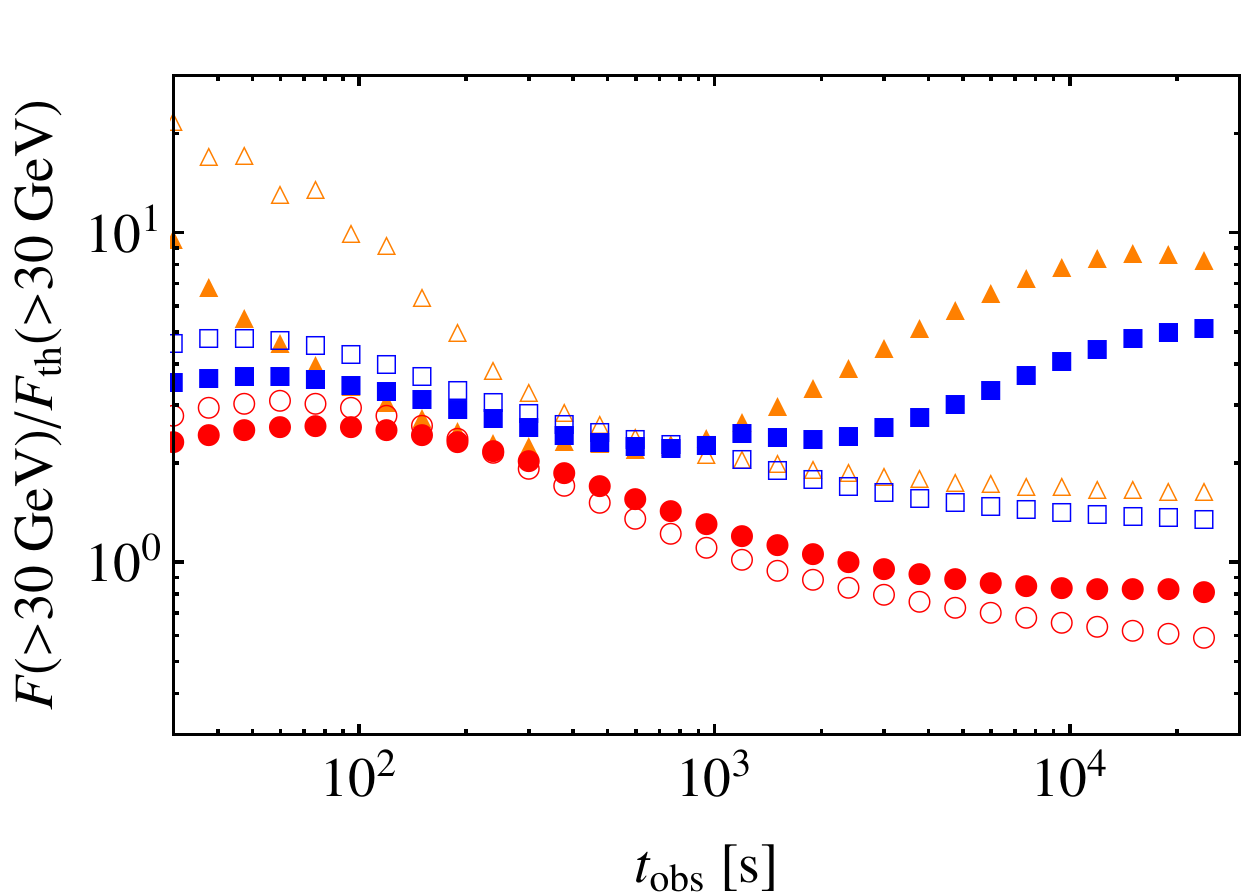}
\caption{Ratio of the total energy flux above $30\,$GeV to a
  theoretical flux obtained by matching the calculated total flux at
  $0.1\,$GeV and extrapolating this flux to higher energies with a
  power-law $F_{\nu,\rm th}\,\propto\,\nu^{-1.1}$, as in Inoue et
  al. (2013). The symbols correspond to the six models studied in
  Figs.~\ref{fig:f2-beta} and \ref{fig:f3-alpha}, as follows: red
  circles $\epsilon_B\,=\,\epsilon_{B+}$ and $\alpha_t\,=\,0$; blue
  squares $\alpha_t\,=\,-0.4$; orange triangles
  $\epsilon_B\,=\,\epsilon_{B-}$ and $\alpha_t\,=\,0$. Filled symbols
  correspond to $k=2$ and open symbols to $k=0$; other parameters are
  as in Fig.~\ref{fig:f1-slow}.
  \label{fig:f4-hilo} }
\end{figure}

One can use the calculations of Sec.~\ref{sec:SSC} to study how a
decaying micro-turbulence or low $\epsilon_{B-}$ affects these
predictions. In order to do so, one calculates the ratio of the total
(synchrotron + inverse Compton) energy flux above $30\,$GeV,
$F(>30\,{\rm GeV})\,=\,\int_{30\,{\rm GeV}} F_{\nu}\,{\rm d}\nu$ to a
theoretical reference flux $F_{\rm th}(>30\,{\rm GeV})$. As in Inoue
et al. (2013), $F_{\rm th}(>30\,{\rm GeV})$ is obtained by
extrapolating the total flux measured at a reference energy, here
$0.1\,$GeV, with an index $\beta\,\simeq\,1.1$, i.e.
\begin{eqnarray}
F_{\rm th}(>30\,{\rm GeV})&\,=\,&F_{\nu}(0.1\,{\rm GeV})\nonumber\\
&& \quad\times\int_{30\,{\rm GeV}/h}^{+\infty}{\rm
  d}\nu\,\left(\frac{h\nu}{0.1\,{\rm GeV}}\right)^{-1.1}
\end{eqnarray}
with $F_{\nu}=F_{\nu,\rm syn}+F_{\nu,\rm IC}$ the total flux.  The
reference energy chosen here is smaller than that in Inoue et
al. (2013), because the inverse Compton flux is already prominent at
$1\,$GeV in the scenarios studied, as shown in Fig.~\ref{fig:f1-slow}
and \ref{fig:f1-fast} for example. However, given the value of the
index $\beta$, the theoretical flux $\nu F_{\nu}$ is roughly flat
above $1\,$GeV, hence this should not affect the statistics of
detection.

The results are shown in Fig.~\ref{fig:f4-hilo}, which carries out
this evaluation for the six models shown in Figs.~\ref{fig:f2-beta}
and \ref{fig:f3-alpha}. This figure indicates that a decaying
micro-turbulence with $\alpha_t=-0.4$ (or a low average $\epsilon_B$)
increases by a factor of a few, up to an order of magnitude, depending
on $\alpha_t$, $k$ and $t_{\rm obs}$, the prospects of observing the
afterglows at energies $>30\,$GeV, relatively to statistics computed
for a model with $\epsilon_B\,=\,0.01$, as in Inoue et
al. (2013). This certainly brings the number of potential detections
by instruments such as CTA in a more confortable range.

Furthermore, it is important to note that in most models studied here,
the inverse Compton component contributes to a significant fraction of
the total flux at $0.1\,$GeV, therefore the above ratio actually is an
underestimate of the ratio of the inverse Compton flux at high
energies to the synchrotron flux at GeV energies.

\section{Conclusions}\label{sec:conc}
This paper has discussed the spectral shapes of the
synchrotron-self-Compton spectrum of a relativistic blast wave,
including all relevant Klein-Nishina effects, and their evolution in
time. A particular emphasis has been put on the impact of a decaying
micro-turbulence behind the shock front, which is motivated by
theoretical analysis (Chang et al. 2008, Lemoine 2015) and
observational inference (Lemoine et al. 2013, Liu et
al. 2013). However, the results are fully applicable to the case of a
uniformly magnetized blast, possibly with a low value of the average
$\epsilon_B$.

A decaying micro-turbulence and/or a low average value of $\epsilon_B$
both lead to a large $Y_{\rm c}$ Compton parameter, with interesting
physical and observational consequences. From a more theoretical point
of view, the Klein-Nishina suppression of inverse Compton cooling has
a strong effect on the synchrotron spectral shape, as noted elsewhere
for the case of a uniformly magnetized blast (Nakar et al. 2009, Wang
et al. 2010). In the case of a decaying micro-turbulence, the
modification is not trivial to compute, and the present paper has
described a simple algorithm which allows to compute the full SSC
spectrum with satisfactory accuracy, at a modest numerical cost. Among
the interesting phenomenological consequences, one may point out the
slight deviations in the spectral and temporal slopes induced by the
decaying micro-turbulence, or by KN effects in uniformly magnetized
blasts at low $\epsilon_B$. A multiwavelength coverage of the
afterglow would, in principle, allow one to tomograph the dynamics of
this magnetized turbulence, through its influence on the light curves
in various wavebands. However, as expressed in terms of the spectral
$\beta$ and temporal $\alpha$ slopes, defined customarily through
$F_\nu\,\propto\,t_{\rm obs}^{-\alpha}\nu^{-\beta}$, the deviations
are relatively weak and not currently distinguishable through
observations. A multiwavelength fit of the afterglow, which also
relies on the flux ratios between various wavebands, seems to provide
a more sensitive probe of the dynamics of the micro-turbulence.

Finally, a large $Y_{\rm c}$ parameter also implies a large inverse
Compton flux at multi-GeV energies, relatively to the lower energy
synchrotron flux, with direct consequences for the detectability of
gamma-ray bursts afterglows by future gamma-ray telescopes. A
numerical estimate indicates that a low average $\epsilon_B$ would
imply a detection rate several times larger than currently anticipated
on the basis of the extrapolation of the flux of gamma-ray bursts
detected by Fermi-LAT.
\bigskip

\noindent {\bf Acknowledgments:} An anonymous referee is acknowledged
for helpful suggestions; H. He is acknowledged for discussions.  This
work has been financially supported by the Programme National Hautes
\'Energies (PNHE) of the C.N.R.S. and by the ANR-14-CE33-0019 MACH
project.

\end{document}